\documentclass[a4paper,11pt]{article}
\pdfoutput=1 

\usepackage{jinstpub} 
\usepackage[flushleft]{threeparttable}
\usepackage[toc,page]{appendix}
\usepackage{booktabs}

\title{Fundamentals of Horn Antennas with Low Cross-polarization Levels for Radioastronomy and Satellite Communications}

\author[a,b,1]{Javier De Miguel-Hern\'andez,\note{Corresponding author.}}
\author[a,b]{Roger J. Hoyland}

\affiliation[a]{Instituto de Astrof\'isica de Canarias,\\E-38200 La Laguna, Tenerife, Spain}
\affiliation[b]{Departamento de Astrof\'isica, Universidad de La Laguna,\\E-38206 La Laguna, Tenerife, Spain}

\emailAdd{jmiguel@iac.es}

\abstract{The literature on horn antennas dedicated to radio astronomy and satellite communications applications is very extensive and at times disjointed, relevant contributions being distributed as far back as from the 60's until the present today.
This work combines a compact but complete review of the different theories, methodologies and techniques used to describe corrugations and metamaterials in their application to feedhorns used in radio astronomy and satellite communications along with some new work to help explain the theory in a more practical way. Starting with the hybrid-mode condition firstly  corrugated horns are explained describing soft and hard boundaries and also the theory from a plasmonic optics point of view. Following this the use of metamaterials in order to design horn antennas with quasi-null cross-polarization and low E-Plane sidelobes level over an ultra-wideband is described.
 The objective of this work is to help to ease the learning curve of the post graduate students and young professionals dedicated to these tasks, and try to inspire the work of the senior professionals toward a new direction and approach.}


\arxivnumber{XXX} 

\begin{document}
\maketitle
\flushbottom

\section{Introduction}
Several characteristics define the performance of a cylindrical horn antenna. Among those treated in the literature are the design of the waveguide and the feedhorn's throat in order to excite only the desired modes, the creation of trapped modes inside the horn because of its internal design, the geometry of the horn and its internal profile and finally, and more relevant to radio astronomy and satellite communications, the geometry of the internal structures needed to support hybrid-modes with low cross-polarization and sidelobe levels. This study aims to summarize these works.

Referring to the last point, the literature on corrugated feedhorn antennas is very widespread and often not well cross-referenced. Different approaches to the theory lead to analogous conclusions, but generally the authors do not focus on the relationship between them.
It is a very difficult task to consult and relate all the references from independent sources, and until now, a simple review which describes and connects all the methods, theories and techniques on the treatment of hybrid-modes to obtain antennas with zero cross-polarization and low E-plane sidelobes level which is optimum for radiastronomy and satellite communications purposes has not existed.

This review contributes by providing a central reference for this topic. In Section \ref{section 2} the origins of the hybrid-mode condition are described. In Section \ref{section 3}, the theory of the corrugated feedhorns is explained. The Section \ref{section 4} is focused on the relation between the classic electromagnetic treatment with the modern plasmonic theory. Section \ref{section 5} looks at hard and soft boundary characteristics and the applications of each. Finally, section \ref{section 6} introduces the state of the art on the use of electromagnetic metamaterials in horn antennas for radioastronomy and satellite applications, which is our ongoing research and section \ref{section 7} shows some examples.

With the idea of allowing a more intuitive and enjoyable read two appendices had been included containing the mathematical derivation of the equations of the electromagnetic field in the case of smooth and corrugated cylindrical waveguides.

\section{The Fundamentals of the Hybrid-mode Condition}
The origins of the {\it hybrid-mode condition} are explained in this section, following through the classic references on this topic.
\label{section 2}

\subsection{The Concept of Surface Impedance}
The concept of surface impedance will be repeatedly used throughout the following section, so a brief definition will be given now.

\begin{figure}[htbp]
		\includegraphics[width=0.4\textwidth]{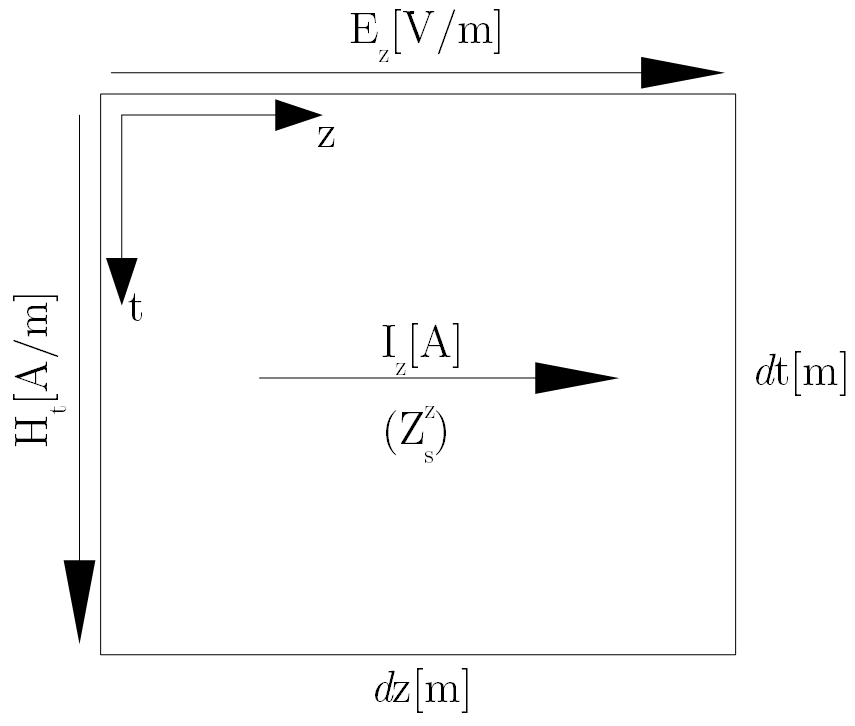}
		\centering
		\caption{Surface impedance schematic.}
		\label{fig1}
	\end{figure}

For the plane surface in the Fig. \ref{fig1} a current along a longitudinal axis (denoted by $\vec{u}_z$) and a transverse axis $\vec{u}_t$ (x or y where $x=r  \, cos \phi$ and $y=r\,sin\phi$ in cylindrical coordinates) exist. Thus, it is possible to define a intensity current and a voltage from a surface wave traveling in the $\vec{u}_z$ direction of this infinitesimal surface and write

\begin{equation}
dI_z[A]=H_t[A/m] \, dt[m]
\label{equation_1}
\end{equation}

\begin{equation}
dV_z[V]=E_z[V/m] \, dz  \; .
\label{equation_2}
\end{equation}

Thus, from the Ohm's Law it is possible to define a surface impedance $Z_s^z$ in the $\vec{u}_z$ direction

\begin{equation}
Z_s^z=\frac {E_z \, dz}{H_t \, dt} = \frac{E_z}{H_t} \;.
\label{equation_3}
\end{equation}

Analogously, an impedance for a wave traveling across the transverse direction of the surface can be defined, leading to $Z_s^t=-E_t/H_z$, where the negative sign exists due to the rotation of the system of coordinates ($t\rightarrow z \, ; \; z\rightarrow -t$) in order to keep the convention of signs, since it implies an inversion of the direction of propagation once the axis are rotated.

\subsection{The Focal-plane Field Reflected by a Paraboloid}

\begin{figure}[htbp]
		\includegraphics[width=0.8\textwidth]{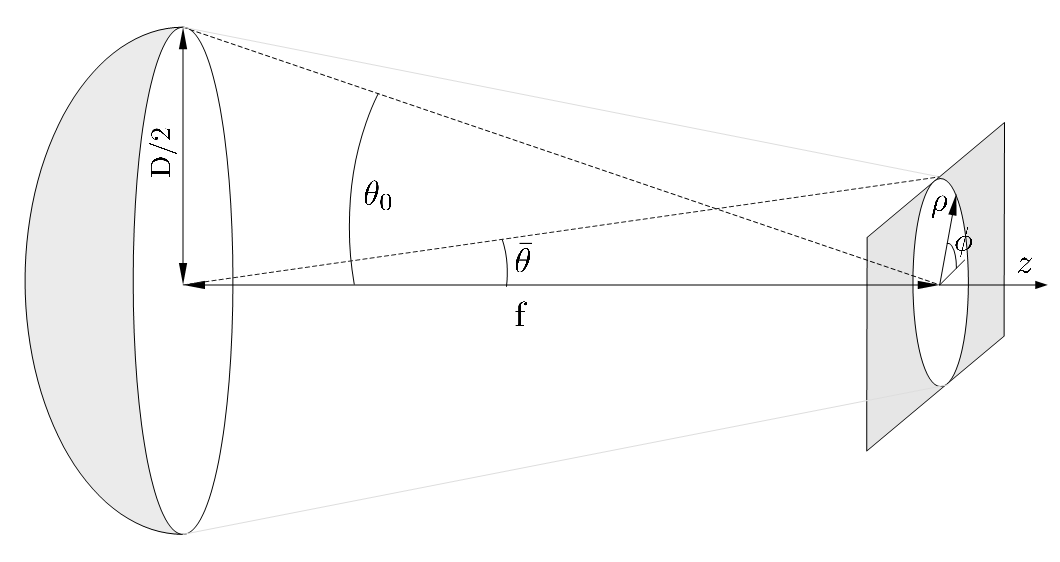}
		\centering 
		\caption{Reflector and focal plane scheme.}
		\label{fig0}
	\end{figure}

Radiotelescopes generally use parabolic reflectors. Fig. \ref{fig0} represents a parabolic reflector and its focal plane. The field generated at the optical focus of a paraboloid with large $f/D$ ratio is formed by the addition of two polarized components \cite{1966ITAP...14..654M}. The transverse magnetic (TM) component is polarized in the plane of incidence and has its electric (E) component oriented normally to the focal plane, generated by an annular ring of the dish subtending a $2\bar{\theta}$ angle at the focus, as shown in the Fig. \ref{fig0} in the focal plane, on the right. The transverse electric (TE) pattern is polarized normal to the plane of incidence and E lies entirely in the focal plane. Since the circles on the focal plane in which the tangential electric field vanishes are not the same for both components, both patterns can't be simultaneously bounded by the same cylindrical conductive tube.

The total field generated by the annular rings of the reflector dish is the combination of the $TM_{1n}$ and $TE_{1n}$ mode patterns, resulting in the hybrid $HE_{1n}$ mode. The expressions of the $HE_{1n}$ fields at a generic point in the focal plane whose coordinates $(\rho,\phi)$ are given by

\begin{equation}
\begin{cases} 
 E_{\rho}=E_{\rho}(\rho)sin\phi\\
 E_{\phi}=E_{\phi}(\rho)cos\phi\\
 E_{z}=E_{z}(\rho)sin\phi\;,
\end{cases} 
\label{equation_4}
\end{equation}

\begin{equation}
\begin{cases} 
 H_{\rho}=H_{\rho}(\rho)cos\phi\\
 H_{\phi}=H_{\phi}(\rho)sin\phi\\
 H_{z}=H_{z}(\rho)cos\phi\;,
\end{cases} 
\label{equation_5}
\end{equation}

where an equal phase term has been assumed  and the $\rho$ dependent field functions are

\begin{equation}
\begin{cases} 
 E_{\rho}(\rho)=-jksin \bar{\theta}\big[cos\bar{\theta}J_0(u)+(1-cos\bar{\theta})\frac{J_1(u)}{u}\big]\\
 E_{\phi}(\rho)=-jksin \bar{\theta}\big[J_0(u)-(1-cos\bar{\theta})\frac{J_1(u)}{u}\big]\\
 E_z(\rho)=-ksin^2\bar{\theta}J_1(u)\;,
\end{cases} 
\label{equation_6}
\end{equation}

where $u=k\rho sin\bar{\theta} \Omega$, $k=2\pi/\lambda$ and

\begin{equation}
\begin{cases} 
 H_{\rho}=-\frac{E_{\rho}(\rho)}{Z_0}\\
 H_{\phi}=\frac{E_{\phi}(\rho)}{Z_0}\\
 H_{z}=-\frac{E_{z}(\rho)}{Z_0}\;,
\end{cases} 
\label{equation_7}
\end{equation}

so the E and H fields of the hybrid mode are identical but rotated expressions.
By reciprocity, a feed which synthesizes the entire focal plane field would radiate a symmetrical pattern producing uniform fields over the dish aperture. The condition for zero cross-polarization for a linear polarized field input requires this symmetry. Appropriate boundaries for hybrid-modes can be developed by considering the circumferential and  longitudinal surface reactances for the feed defined as

\begin{equation}
\begin{cases} 
 X_{\phi}=-j\frac{E_{\phi}}{H_z}\\
 X_{z}=j\frac{E_{z}}{H_{\phi}}\;,
\end{cases} 
\label{equation_8}
\end{equation}

Inserting (\ref{equation_6}) and (\ref{equation_7}) into (\ref{equation_8}) leads to the following condition for the required surface boundary of the waveguide

\begin{equation}
X_{\phi}X_z=-Z_0^2 \;,
\label{equation_9}
\end{equation}

Thus, a waveguide with a inner surface which satisfies (\ref{equation_9}) supports the hybrid-mode and receives the complete field pattern from a paraboloid with null cross-polarization. The relation in (\ref{equation_9}) is known as {\it hybrid-mode condition}. This is a necessary but not sufficient condition to attenuate cross-polarization. Other factors such as the profile of the antenna or the antenna-reflector array optics also influence the final levels of cross polarization.

\section{The Classic Corrugated Conical Feed-horn}
\label{section 3}

\citet{1966ITAP...14..654M} introduce  the use of radially corrugated cylindrical waveguides to support the hybrid-mode condition  (\ref{equation_9}) in their paper. Other authors make a more complete analytical treatment of this technology, which is the preferred choice for radioastronomy and satellite communications horn antennas even today. \newline
There are many published contributions to feedhorn design centred over a complete range of frequencies. They vary in how the horn is optimized. It depends on the design criterion. A horn can be optimized for return-loss, gain, sidelobe level, cross-polar response or other mechanical consideration (size, mass, manufacturability).

\subsection{The Corrugated Feedhorn optimized for Return Loss}

The Return Loss of the feedhorn depends almost entirely on the throat section design which has the job of matching the waveguide fundamental mode $TE_{11}$ to the corrugated cylinder mode $HE_{11}$. There are published design parameters for smooth to corrugated wall waveguides \cite{2005ATM}. In this reference, there are 3 basic designs with corresponding Return Loss bandwidths, the widest bandwidth being achieved by using a ring-loaded corrugation design.

\subsection{The Corrugated Feedhorn Optimized for Gain or Directivity}

The gain or directivity of the corrugated feedhorn is a function of its aperture size. There are other factors to take into account such as the edge taper (for soft surfaces) and the horn flare angle to reach the given aperture. The profile of a corrugated feedhorn will also affect its gain. Design parameters are given in \cite{2005ATM} and commercial programs such as MAGNUS$^{\textregistered}$. 
\subsection{The profiled Corrugated Feedhorn }

Although the introduction of a profile may result in a shorter structure, it does come at the expense of raised side lobe levels and higher cross-polarization, mainly due to the $HE_{12}$ mode excited by the varying flare angle along the profile.

Each profile performs quite differently, some better than the standard linear profile, and some worse. Typically, the sinusoidal and polynomial profiles provide the best compact alternative to the linear profile.

Comparing structures designed for equal gain, it is clear how a correct profile results in a more compact structure, albeit at the expense of increased side lobes. Also, in the case where the profiled structure is comparative in size to the linear one (see hyperbolic vs linear) the overall performance of the profiled version seems slightly better, with lower shoulders than the linear version.

\subsection{The Corrugated Feedhorn Optimized for Lowest Sidelobe Level}

 An important factor for Radio Astronomy or Telecommunications is the sidelobe level for an antenna. Corrugated feedhorns can achieve very low sidelobes  \cite{2002MWC} using a Gaussian profile. The low sidelobe level is due to the ability of the feedhorn shape to convert the $HE_{11}$ mode into a gaussian cross-section across the aperture. This can be achieved by converting a small percentage of the $HE_{11}$ mode into higher order modes until the mix is such that a Gaussian cross-section is achieved.  \newline
 The sidelobe level is often grouped with cross polarization performance in a misleading way. Good sidelobe level does not lead to good cross-polarization. Indeed, a horn with high sidelobe levels can still be designed to have low cross polarization. The condition for the lowest cross-polarization is described in the following sections and relates to a pure $HE_{11}$ mode. Mixing in other higher order modes will change the  theoretical zero cross-polar condition.
 
 Recent research suggests that Gaussian profiled horns can contribute to a pure gaussian-mode composed by several $HE_1n$ modes, leading to very low side-lobes levels too \cite{Del Río}.

\subsection{The Corrugated Feedhorn and the Hybrid-mode Condition}

A smooth metallic cylindrical waveguide cannot satisfy the hybrid-mode condition (\ref{equation_9}). This is shown in the Appendix-\ref{Appendix-A} of this article. The first surface solution found to satisfy the hybrid condition was the corrugated cylindrical or conical feed-horn. Other solutions involve profiled smooth-walled cylindrical feedhorns which make use of concentric steps or tapers to convert and phase align TM modes to equivalent TE modes. These are limited in cross-polar isolation and bandwidth \cite{Zeng}.

\begin{figure}[ht!]
		\includegraphics[width=0.65\textwidth]{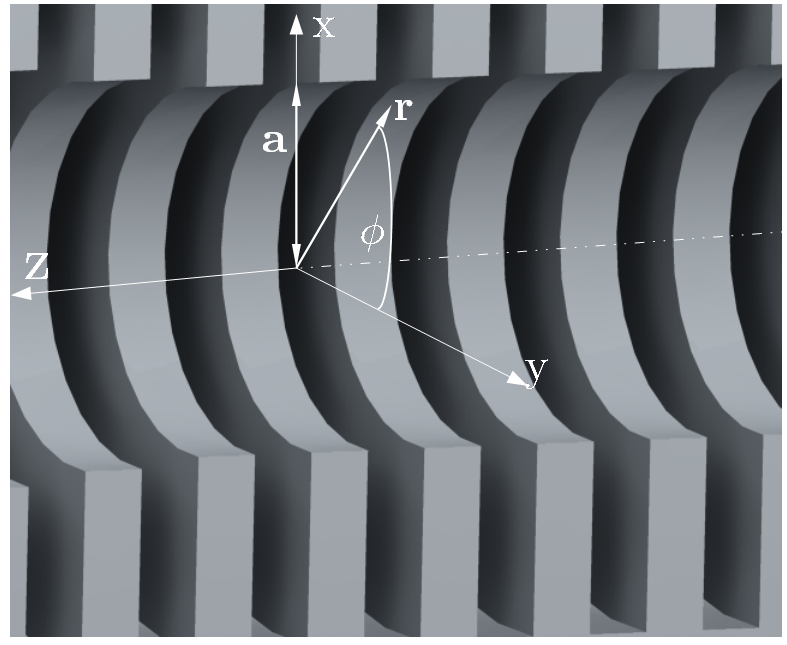}
		\centering 
		\caption{Corrugated cylindrical horn. 3D xz-cut.}
		\label{fig2}
	\end{figure} 

In this section, the use of cylindrical corrugations in a waveguide with circular section (see Fig. \ref{fig2}) is introduced with the aim of supporting and correctly transmitting the hybrid-mode (approximately composed of  84.5\% of $TE_{11}$,  14.6\% of $TM_{11}$ and a small,  $\le{ 1\%}$, contribution of other modes). Since (soft surface (section 5.1)) corrugated horns present a symmetric and tapered aperture distribution these characteristics provide farfield patterns with low sidelobes as well as low cross polarization. The method given by \citet{Clarricoats} and \citet{Dragone} for microwave antennas is followed here, in order to preserve the historical development of the theory. Other important points in the design of feedhorns with low cross polarization, such as higher order modes produced in the throat-region, the higher order mode conversion along the length of the horn and direct radiation from the flange (only for small aperture horns) which are not the objective of this work, can be also consulted in these references. In general it is the intrinsic cross polarization of the corrugated structure that dominates the overall cross polarization, however, when this has been reduced to a sufficiently low level the other terms become significant.\newline

For a smooth waveguide, an expression for the propagation constant $\beta$ can be relatively easily obtained, as can be seen in the Appendix- \ref{Appendix-A} of this paper, which is based on work by \citet{Pozar}. The expression can be found in equations (\ref{equation_A27}) for the case of TM modes and (\ref{equation_A37}) for TE modes. Here it can be seen that the $TE_{1\, 1}$ is the fundamental mode of a cylindrical waveguide so, in a frequency range above the cut-off, it can be theoretically excited in the absence of higher modes . The $TM_{0 \, 1}$ is the next propagating mode of a cylindrical waveguide. Also shown in Appendix \ref{Appendix-A} is the fact that a smooth-walled metal cylinder cannot satisfy the hybrid mode condition since the surface impedance only depends on the properties of the metal. 
\newline

\begin{figure}[ht!]
		\includegraphics[width=0.8\textwidth]{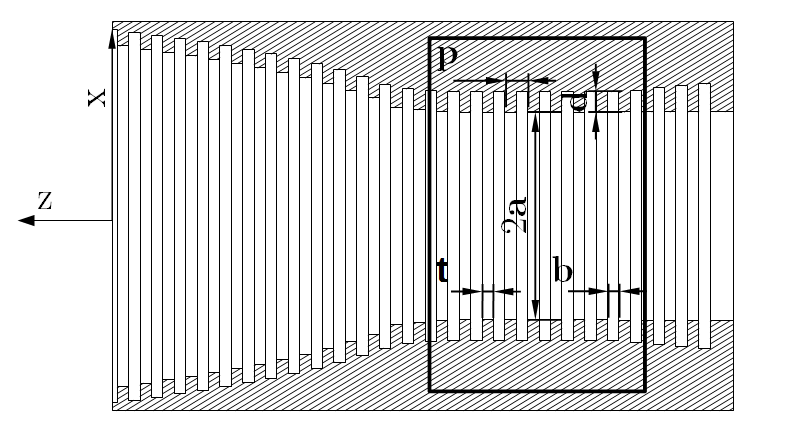}
		\centering 
		\caption{Corrugated horn section. The inner box highlights the studied zone. }
		\label{fig3}
	\end{figure}

For the case of a practical corrugated horn shown in Figs. \ref{fig2} and \ref{fig3}, a simple relation for the propagation function cannot be found for the general case. An important exception is the case of a horn with a big aperture radius ($a$) compared to the wavelength ($\lambda$). Under this condition, two expressions are enough to describe the propagation function of all the modes except one\footnote{Consult Dragone Appendix-B.}. The derivation of these propagation functions ($\beta_{nl}$) is shown in the Appendix-\ref{Appendix-B} of this paper, and the propagation functions are given by (\ref{equation_B1}) and (\ref{equation_B2}).

The field expressions for transverse waves in the corrugated waveguide are derived from (\ref{equation_B10}) to (\ref{equation_B15}), leading to \newline

\begin{equation}
\vec{E_t}=-j\frac{k\,a}{u}A\left[J_0\left(\frac{r}{a}u\right)\vec{u}_x+\frac{1}{4}u^2\frac{y}{k\,a}J_2\left(\frac{r}{a}u\right)\Big(cos(2\phi)\vec{u}_x+sin(2\phi)\vec{u}_y\Big)\right]e^{-j\beta z} \;,
\label{equation_39}
\end{equation}

where $t$  refers to whatever transverse direction $x$ or $y$ in  Fig. \ref{fig2} and for $HE_{1\,l}$ modes, and 

\begin{equation}
\vec{E_t}=-j\frac{k\,a}{u}A\left[J_2\left(\frac{r}{a}u\right)\Big(cos(2\phi)\vec{u}_x+sin(2\phi)\vec{u}_y\Big)\right]e^{-j\beta z} \;,
\label{equation_40}
\end{equation}

for $HE'_{1\,l}$ modes. \newline

The perfect response for a telecommunications or astrophysical horn is given by a Gaussian radial distribution at the aperture and with a smooth finite edge taper, symmetric azimuthal distribution and a single linearly polarized direction for the E field, and this can only be obtained in the limit $y/ka$ for the $HE_{1\,l}$ modes, so the $HE'_{1\,l}$ modes must be avoided because they contain a cross-polarized component whose amplitude is independent of the ratio $y/ka$ (\citet{Dragone2}, pp. 869-888). For this reason, the only way to achieve the desired response is to not excite them at all or cut them off in the \textit{launching region} of the horn. This can be done over a relatively wide frequency band, converting almost all the power incident on the input of the feedhorn into the fundamental $HE_{11}$ mode.\newline
\newline Given that the $HE'_{1\,l}$ modes can be avoided an important consequence can be obtained from (\ref{equation_39}). That is 

\begin{equation}
k\,a\rightarrow\infty \  	\Rightarrow E_y\rightarrow 0  \;.
\label{equation_41} 
\end{equation} \newline

Thus, in the limit $k\,a \rightarrow \infty$, the field becomes polarized in one direction independently of the value of the surface reactance $X_s$ (remember $y=-Z_0/X_s$) unless $X_s=0$. Thus, in order to obtain $E_y\simeq 0$ over a specific band, it is enough to restrict the feed to having a large aperture radius (i.e., large $a$) and a corrugation thickness small in comparison with the corrugation separation (i.e., $t<<b$)\footnote{Note that this is not the case in Fig. \ref{fig3}, where a different configuration has been adopted for others reasons e.g. ease of manufacture.}.

In addition to this, since the second term in (\ref{equation_39}) introduces beam asymmetry and cross-polarization and the aperture is physically limited (so the $ka$ product is limited to $k\,a < \infty$), it is necessary to make $y\rightarrow 0$, which in practice implies a corrugation depth of $d\rightarrow \lambda/4$. 

In real horn antennas, $d\simeq \lambda/4$ is satisfied at the center of the operation band by correctly choosing the corrugation depth, and large $ka$ is achieved near the horn aperture. The reason for the condition $d\simeq\lambda/4 \Rightarrow X_s\rightarrow\infty$ will be explained in section \ref{section 3.2}.\newline
\newline In Clarricoats et al. another more generalized form of (\ref{equation_39}) is given. This is

\begin{equation}
\vec{E_t}=-j\frac{k\,a}{u}A\left[J_0\left(\frac{r}{a}u\right)\vec{u}_x+\frac{1}{4}u^2\frac{X-Y}{k\,a}J_2\left(\frac{r}{a}u\right)\Big(cos(2\phi)\vec{u}_x+sin(2\phi)\vec{u}_y\Big)\right]e^{-j\beta z} \;,
\label{equation_42}
\end{equation}

where $X$ is the surface reactance in the transverse direction in cylindrical coordinates ($\phi$), and its expression, which can be derived from the definition of surface impedance in (\ref{equation_3}), is given by

\begin{equation}
X=-j\frac{Z_{\phi}}{Z_0}=-j\frac{E_{\phi}}{H_z}\sqrt{\varepsilon_0/\mu_0} \;,
\label{equation_43}
\end{equation}

and $Y$ is the admitance in the direction of propagation ($z$) and is given by

\begin{equation}
Y=j\frac{Z_0}{Z_{z}}=j\frac{H_{\phi}}{E_z}\sqrt{\mu_0/\varepsilon_0} \;.
\label{equation_44}
\end{equation}

Since for a set of concentric metallic corrugations $E_{\phi}\simeq 0$ due to the low value of $Z_{\phi}$, because there are not discontinuities but only good conducting metallic paths for the surface waves in the transverse direction, it is possible to derive (\ref{equation_39}) from (\ref{equation_42}), (\ref{equation_43}) and (\ref{equation_44}) remembering that $Z_0=\sqrt{\mu_0/\varepsilon_0}$. The following equations apply for the corrugated scheme

\begin{equation}
y=X-Y= -j\frac{E_{\phi}}{H_z}\sqrt{\varepsilon_0/\mu_0} - j\frac{H_{\phi}}{E_z}\sqrt{\mu_0/\varepsilon_0} \simeq -j\frac{H_{\phi}}{E_z} Z_0  = -j\frac{Z_0}{X_s^z}\;,
\label{equation_45}
\end{equation}

which recovers the definition of $y$ given by (\ref{equation_B17}).\newline

Thus, (\ref{equation_42}) establishes that $X-Y=0$  in order to maintain the hybrid-mode condition. Remembering that the TE modes have $E_y$ polarization ({\it i.e.}, $Z_s^y \equiv Z_s^{TE}$) and the TM modes present $E_z$ polarization ($Z_s^x \equiv Z_s^{TM}$), the $X-Y=0$ condition can be expressed in terms of surface impedance of the corrugated horn in both propagation and transverse directions ($z, \,\phi$ respectively). The expression is

\begin{equation}
X-Y=0\rightarrow -j\frac{E_{\phi}}{H_z}Z_0^{-1} - j\frac{H_{\phi}}{E_z}Z_0 =0 \;,
\label{equation_46}
\end{equation}

and by substitution of the surface impedances here the following equation is obtained

\begin{equation}
\frac{Z_0}{Z^{TM}}+\frac{Z^{TE}}{Z_0}=0\;.
\label{equation_47}
\end{equation}

Since in a good metallic conductor the resistance is near zero ($Z=R+jX\simeq jX$) equation  (\ref{equation_47}) reduces to

\begin{equation}
\frac{Z_0}{X^{TM}}+\frac{X^{TE}}{Z_0}=0\;.
\label{equation_48}
\end{equation}

By multiplying (\ref{equation_48}) by $Z_0$ and reordering terms, the condition for maintaining the hybrid-mode is obtained again

\begin{equation}
X^{TE}\, X^{TM}=-Z_0^2\;.
\label{equation_49}
\end{equation}

It's clear that (\ref{equation_9}) and (\ref{equation_49}) are equivalent expressions, so a corrugated feedhorn satisfying $X-Y=0$ transmits all the field in the focal-plane from a parabolic reflector with null cross-polarization.

\subsection{The Transmission Line and Parallel-LC Equivalent Circuits} \label{section 3.2}

Two different approximations can be taken in order to explain the behaviour of surface waves travelling along a corrugated wall. In this section, these two equivalent circuits are presented.

\subsubsection{The Transmission Line equivalent circuit}
Observing the array of grooves in  Fig. \ref{fig3} an approximate equivalent circuit can be defined, where each corrugation can be described by a transmission line where an incident wave passes along the line until finding a discontinuity at the end ($x=-d$) which is the bottom of the groove.

\begin{figure}[ht!]
		\includegraphics[width=0.4\textwidth]{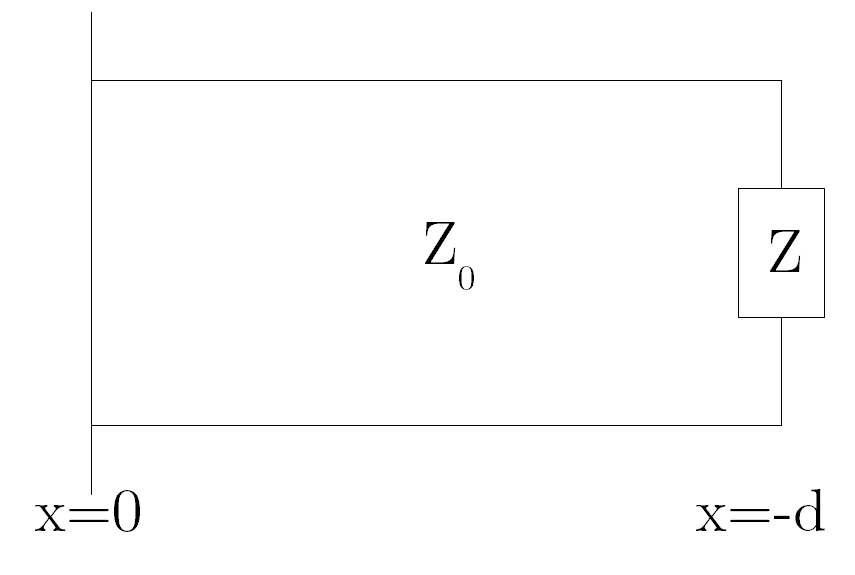}
		\centering 
		\caption{Corrugation equivalent circuit. }
		\label{fig4}
	\end{figure}

When $d\simeq\lambda/4$ the short circuit ($\left| Z_s\right| \rightarrow0$) at $x=-d$ is transformed into an open circuit ($\left| Z_s\right| \rightarrow\infty)$ at the top ($x=0$). 
This equivalent circuit is represented in Fig. \ref{fig4}, where the load represents the surface at the bottom of the groove.

As described in \citet{Sievenpiper}, due to the reflection from the bottom of the groove a standing wave will be generated by the composition of an incident and a reflected wave. The wave can be described by (\ref{equation_50}) and (\ref{equation_51}) 

\begin{equation}
E_x=E^i\, e^{jkx}+E^r\,e^{-jkx}  \;
\label{equation_50} 
\end{equation}

\begin{equation}
H_x=H^i\, e^{jkx}+H^r\,e^{-jkx}  \;.
\label{equation_51} 
\end{equation}

At the bottom of the groove ($x=-d$) the boundary condition is

\begin{equation}
\frac{E_{x=-d}}{H_{x=-d}}=Z  \;,
\label{equation_52} 
\end{equation}

and in general the H and E fields are related by the characteristic impedance of the transmission line in the form

\begin{equation}
\left | \frac{ E_{x}^i}{H_{x}^i}\right|= \left | \frac{ E_{x}^r}{H_{x}^r}\right| = Z_0\;.
\label{equation_53} 
\end{equation}

Since both electric and magnetic fields are reversed by the reflection, it is possible to write

\begin{equation}
E_{x=-d}^i=-E_{x=-d}^r\;.
\label{equation_54} 
\end{equation}

The combination of (\ref{equation_50}), (\ref{equation_51}), (\ref{equation_53}) and (\ref{equation_54}) leads to the expression of the impedance as a function of the distance (i.e., $x$) from the short circuit

\begin{equation}
\frac{E_{x}}{H_{x}}=\frac{E^i\, e^{jkx}+E^r\,e^{-jkx}}{H^i\, e^{jkx}+H^r\,e^{-jkx}}=\frac{E^i\, e^{jkx}-E^i\,e^{-jkx}}{H^r\, e^{jkx}-H^r\,e^{-jkx} }  \;.
\label{equation_55} 
\end{equation}

At the surface on the top of the corrugation ($x=0$), it is possible to write

\begin{equation}
Z_{x=0}=j\, Z_0\, tan(k\,d)\;,
\label{equation_56} 
\end{equation}

where $k=2\pi/\lambda$ is the wave number. 
The analysis of (\ref{equation_56}) leads to

\begin{equation}
Z_{x=0}
\begin{cases} 
 \le 0 \;\;\;  \mbox{if } \,  \lambda/2 > d > \lambda/4 \\
 >0 \;\;\;  \mbox{if } \, d< \lambda/4  \\ 
 =\infty \;\; \mbox{if } \, d=\lambda/4\,.
\end{cases} 
\label{equation_57}
\end{equation} \newline

For the reference system in Fig. \ref{fig3}, and for the case of a transverse magnetic (TM) surface wave, the boundary condition is $H_x=H_z=E_y=0$, so the propagation is given by

\begin{equation}
E_z=C\,e^{-j(k_zz-k_xx+\omega t)}  \;,
\label{equation_58}
\end{equation}

where C is a time independent arbitrary constant. From Maxwell's Equations (in the absence of stored charges) it is possible to write

\begin{equation}
j\, \omega \varepsilon E_z=\frac{\partial H_y}{\partial x} \rightarrow H_y=-j\frac{\omega \,\varepsilon}{k_x}C\,e^{-j(k_zz-k_xx+\omega t)}\;.
\label{equation_59}
\end{equation}

By dividing (\ref{equation_58}) and (\ref{equation_59}) the following equation can be  obtained

\begin{equation}
Z_s^z\equiv Z_s^{TM}=\frac{E_z}{H_y}= j\frac{k_x}{\omega \varepsilon}    \;.
\label{equation_60}
\end{equation} \newline

A generic impedance is defined as $Z=R+jX$, where $R$ is the resistance and the reactance $X$ has inductive and capacitive components, {\it i.e.}, $X=X_L+X_C=\omega \,L-\frac{1}{\omega \, C}$. From (\ref{equation_60}), since $Z_s^{TM}$ is purely imaginary and positive, it can be derived that the impedance which supports TM waves is inductive ($X_s^{TM}=X_L$). Since it is well known that a pure inductance establishes a $+\pi/2$ phase-shift\footnote{An impedance is defined as $Z=R+jX \equiv |Z|e^{j arg(Z)}$ where $arg(Z)$ represents the phase-shift between voltage and current, this is $+\pi/2$ rad for inductors and $-\pi/2 $ rad for capacitors.}, this phase-shift is expected to be seen in the measurement of the reflection.\newline

The analogous analysis can be done for a TE surface wave. The TE boundary conditions are $H_y=E_z=E_x=0$ so

\begin{equation}
H_z=C\,e^{-j(k_zz-k_xx+\omega t)}  \;,
\label{equation_61}
\end{equation}

and from Ampere's Circuital Law the following can be obtained

\begin{equation}
\vec{\nabla} \times \vec{E} = - \frac{\partial{\vec{B}}}{\partial{t}} =-\mu \frac{\partial H}{\partial{t}} \;,
\label{equation_62}
\end{equation}

and since $Z_s^y \equiv Z_s^{TE}= -\frac{E_y}{H_z}$ (where the negative sign can be understood as a rotation of the coordinate system around the $x$ axis) finally the following can be obtained

\begin{equation}
Z_s^y\equiv Z_s^{TE}= -j\frac {\omega \mu}{k_x}   \;.
\label{equation_63}
\end{equation}

From (\ref{equation_63}), and since $Z_s^{TE}$ is purely imaginary and negative, it can be deduced that the impedance which supports TE waves is capacitive ($X_s^{TE}=X_C$). Since it is well known that a pure inductance establishes a $-\pi/2$ phase-shift, this phase-shift is expected to be seen in the measurement of the reflection.\newline

Therefore, only surfaces with pure inductive reactance can support TM modes, while only capacitive surfaces can support TE modes. Between both extremes, other possibilities exist.\newline
\subsubsection{The equivalent LC circuit analysis}

\begin{figure}[ht!]
		\includegraphics[width=0.5\textwidth]{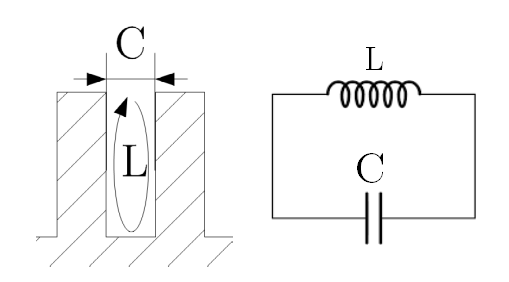}
		\centering 
		\caption{Corrugations (left) equivalent LC parallel circuit (right). }
		\label{fig00}
	\end{figure}

An LC equivalent circuit can be established. Here the capacitive part of the surface reactance $X_s$ ($X_C$) is a result of the wave passing through two parallel plates (corrugation walls) while the inductive part, $X_L$ is a result of the circulation of the wave around the two parallel plates which are shorted at one end. This is represented in Fig. \ref{fig00}. It's important to emphasize that this physical model is a simplification of what really happens.\newline

It is well known that in LC parallel circuits a resonant frequency $\omega_0$ exists at which the inductive and capacitive reactances are equal in magnitude ($X_{C}=X_{L}\,\!$) . Therefore, at the resonance frequency the impedance will be minimal and equal to the ohmic resistance. The following equations (\ref{equation_64}) and (\ref{equation_65}) express this

\begin{equation}
\omega _{0}={\frac  {1}{{\sqrt  {LC}}}} \;
\label{equation_64}
\end{equation}

\begin{equation}
\frac{1}{Z}={\sqrt {\frac{1}{R^{2}}+\left(\frac{1}{X_{C}}-\frac{1}{X_{L}}\right)^{2}}} \;,
\label{equation_65}
\end{equation} \newline

It can be seen that in (\ref{equation_65}) if $X_L = X_C$ then $Z$ = $R$. It can also be said that a limiting value of $Z$ exists, given by the impedance of free space, which is a pure real resistance because in the vacuum there are no distributed charges ({\it i.e.}, $Z_0=119.9169832\pi \,  \Omega$). If $Z_0$ is surpassed, the surface wave is liberated into the medium with lower impedance. Thus, surface waves exist, when $\omega\rightarrow \omega_0 \Rightarrow Z_s\rightarrow \infty$, until a limiting value $Z_s\simeq 377 \,  \Omega$.

In this LC-model, the surface impedance of the corrugated-horn is given by

\begin{equation}
Z_s=\frac{j\omega L}{1-\omega^2LC} \;.
\label{equation_66}
\end{equation} \newline

Since its expression is $\Phi=arctan[(X_L-X_C)/R]=0$, when $\omega=\omega_0$ the phase-shift is exactly $\pm 0 \,$ rad. A phase-shift of $\pm \pi/2 \,$ rad is given by purely inductive or capacitive reactances. This LC equivalent circuit describes the reality satisfactorily in the vicinity of the resonant frequency, {\it i.e.}, $\omega_0=2\pi f_0$.

For a surface TM wave, as in (\ref{equation_58}) and by employing anew the Ampere's and Faraday's laws the following is derived

\begin{equation}
j\omega\varepsilon_0 E_z=\frac{\partial H_y}{\partial x} \;
\label{equation_67}
\end{equation}

\begin{equation}
j\omega\varepsilon_0 E_x=\frac{\partial H_y}{\partial z} \;
\label{equation_68}
\end{equation}

and 

\begin{equation}
-j\omega \mu_0 H_y=\frac{\partial E_x}{\partial z} - \frac{\partial E_z}{\partial x} \;,
\label{equation_69}
\end{equation}

resulting in

\begin{equation}
E_x=-j \frac{\omega k_z}{k_x} C e^{-j(k_zz-k_xx+\omega t)}\;.
\label{equation_70}
\end{equation}

The combination of (\ref{equation_59}) and (\ref{equation_70}) leads to the following expression of the wavevector in the $\vec{u}_z$ direction

\begin{equation}
k_z^2=\mu_0\varepsilon_0\omega^2+k_x^2\;,
\label{equation_71}
\end{equation}

and (\ref{equation_60}) and (\ref{equation_71}) lead, after some manipulation, to

\begin{equation}
k_z^2\equiv k_{TM}^2= \frac{\omega^2}{c^2}   \left(1-\frac{Z_s^2}{Z_0^2}\right )   \;,
\label{equation_72}
\end{equation}

where $Z_0=\sqrt{\mu_0/\varepsilon_0}$. \newline

The analogous process for TE waves leads to the following expression for the wave vector

\begin{equation}
k_y^2\equiv k_{TE}^2= \frac{\omega^2}{c^2}   \left(1-\frac{Z_0^2}{Z_s^2}\right )   \;,
\label{equation_73}
\end{equation}

where $c$ is the speed of light. \newline

Assuming a lossless case, from (\ref{equation_72}) it can be deduced that the propagation of TM surface waves is only possible when $Z_s^{TM} < Z_0$. 
From (\ref{equation_73}) it can be deduced that the propagation of TE surface waves is only possible when $Z_s^{TE} > Z_0$. 
On the other hand, from (\ref{equation_52}), (\ref{equation_53}) and (\ref{equation_55}) it is possible to derive  the phase difference between the incident and the reflected waves

\begin{equation}
\Phi =   Im \Bigg\{ln \left(\frac{E^r}{E^i}\right) \Bigg\}=Im \Bigg\{ln \left(\frac{Z_s-Z_0}{Z_s+Z_0}\right) \Bigg\}  \;,
\label{equation_74}
\end{equation}

where $Z_s$ is given by (\ref{equation_66}). \newline

\begin{figure}[ht!]
		\includegraphics[width=0.8\textwidth]{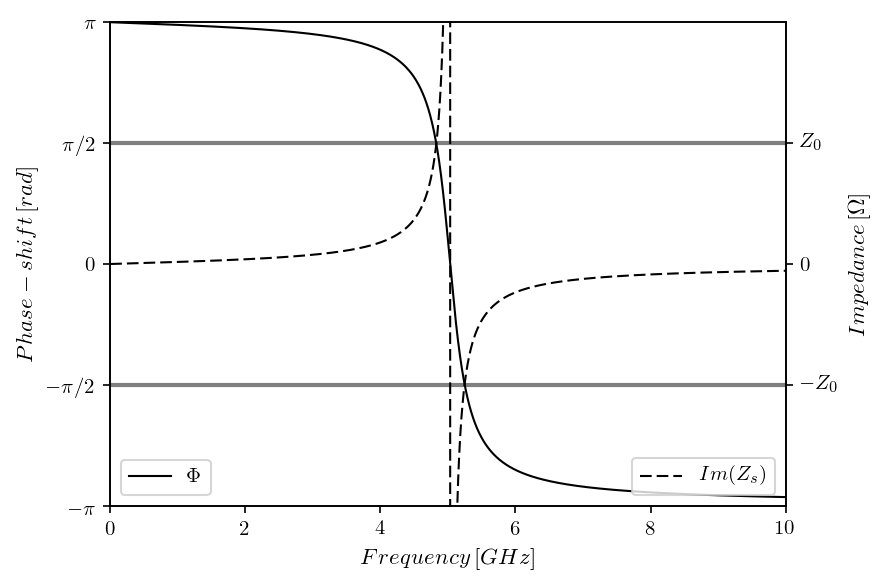}
		\centering 
		\caption{Phase-shift($\Phi$) and surface impedance ($Z_s$) relation for a surface with high impedance for $L=1$ $nH^2$ and $C=1$ $pF^2$.}
		\label{fig000}
	\end{figure}

From (\ref{equation_74}) it can be deduced that when $Z_s$ in a certain direction is low, the reflection phase ($\Phi$) of an incident wave is $\pm \pi$. While for increasing $Z_s$  the value of $\Phi$ tends to zero ($\Phi\rightarrow 0$), and the critical value of $Z_s$ being the medium impedance ($Z_s\rightarrow Z_0$). This is represented in the Fig.\ref{fig000} where the phase shift and surface impedance is plotted together. If we accept that the surface waves are coupled to the corrugated surface only when $|Z_s|<Z_0$ then this will only occur when the surface causes small phase shifts coinciding exactly with the $[-\pi/2,\pi/2]$ phase-shift interval.

Notice that the conclusions extracted from the previous analysis of the equivalent LC-circuit in (\ref{equation_65}) and from the reflection of a wave in (\ref{equation_74}) are analogous. For the LC-circuit it was deduced that for purely inductive surfaces ($X_s=X_L$) a phase-shift of $+\pi/2$ is expected, while for purely capacitive surfaces ($X_s=X_C$) it is expected a phase-shift of $-\pi/2$.
Only at the resonance frequency, the match of reactances is reached ($X_L=X_C$) and the surface impedance $Z_s$ is a purely real number ($Z_s=R=Z_0$).

\subsection{The Standing Waves Analysis}

The possibility of studying the standing waves (SW) generated inside the grooves is presented in the following text. This concept has been briefly mentioned before, but here a more detailed treatment will be explored.\newline
\newline For the guide in Fig. \ref{fig3} it has been  already established that a TM surface wave travelling in the $\vec{u}_z$ direction will generate a standing wave inside the groove and between parallel plates in the $\vec{u}_x$ direction. 

To represent the SW effect we assume that incident or forward plane wave is travelling through the $x$ axis, while a reflected or backward plane wave is travelling on the $-\vec{u}_x$ direction. If we assume a lossless medium the waves can be represented as

\begin{equation}
A_{1}(x,t)=A_0sin\left(\frac{2\pi x}{\lambda}-\omega t \right)
\label{equation_75}
\end{equation}

\begin{equation}
A_{2}(x,t)=A_0sin\left(\frac{2\pi x}{\lambda}+\omega t \right) \; ,
\label{equation_76}
\end{equation}

where $k=2\pi/\lambda$. The combination of both waves generate a standing wave inside the groove with the form

\begin{equation}
A(x,t)=A_{1}(x,t)+A_{2}(x,t)=A_0sin\left(\frac{2\pi x}{\lambda}-\omega t \right) + A_0sin\left(\frac{2\pi x}{\lambda}+\omega t \right) \;.
\label{equation_77}
\end{equation}

From (\ref{equation_77}), by using the {\it sum-to-product} trigonometrical identity and reordering terms we obtain

\begin{equation}
A(x,t)=2A_0sin\left(\frac{2\pi x}{\lambda}\right) cos(\omega t) \;,
\label{equation_78}
\end{equation}

where the $sin$ term of the equation implies the existence of even multiples of $\lambda/4$ called \textit{nodes} at the points where the value $A(x,t)=0$, and odd multiples of $\lambda/4$, called \textit{anti-nodes}, where the amplitude is maximum.

Thus, if a groove depth $x=d=\lambda/4$ is chosen in Fig. \ref{fig3} an anti-node is established exactly at the top of the groove and a node at the bottom of the groove. Hence, a virtual short circuit is generated at the bottom of the groove, while a virtual open circuit is generated at the top. Since the impedance of an open circuit is infinite, the condition $Z_s^{TM}\rightarrow \infty$ is reached when $d \rightarrow  \lambda/4$.\newline

On the other hand, this  standing wave generated in the $x$ direction will be attenuated while it is travelling along the $x$ axis from the top of the groove to the open space\footnote{This concept is related with the exponential $e^{-1}$ decay \textit{skin depth} $\delta_s=\frac{1}{k_{x_j}}=\sqrt{\frac{2}{\omega \mu\sigma}}$, where $\sigma$ is the conductivity.}. Its attenuation constant ($\alpha$) is related with the propagation constant in this direction ($\gamma$) by the following relation

\begin{equation}
\frac{A_0}{A_x}=e^{\gamma x} \;,
\label{equation_79}
\end{equation}

where $\gamma=\alpha+j\beta$, $\alpha$ is the attenuation constant and $\beta$ is the phase constant.
Thus

\begin{equation}
\bigg|\frac{A_0}{A_x}\bigg|=e^{\alpha x} \;.
\label{equation_80}
\end{equation}

If the possibility of wave penetration into the metallic plate on the bottom of the groove is considered, it could be analogously possible to define an equivalent attenuation or decay constant along the negative $x$ axis, here called $\alpha'$, which gives an indication of the penetration depth of the wave into the metal.

By a similar procedure we used to obtain (\ref{equation_58}) and by using Maxwell's Equations, it can be derived that the expressions for both attenuation constants are 

\begin{equation}
E_{x>0}=C \, e^{-k_z z-\alpha x} 
\label{equation_81}
\end{equation}

\begin{equation}
E_{x<0}=C \, e^{-k_z z-\alpha' x} \; ,
\label{equation_82}
\end{equation}

where the term $e^{j\omega t}$ is implicit, and since

\begin{equation}
\vec{\nabla} \times \vec{B} = \frac{\varepsilon_r}{c^2} \frac{\partial{\vec{E}}}{\partial{t}}  \;
\label{equation_83}
\end{equation}

it is possible to derive 

\begin{equation}
\vec{\nabla} \times \vec{\nabla} \times \vec{E} = -\frac{\varepsilon_r \omega^2}{c^2} \frac{\partial^2{\vec{E}}}{\partial{t}^2}  \;.
\label{equation_84}
\end{equation}

Since it is known that $E_y=0$ because the analysis here is considering a surface wave in the $\vec{u}_z$ direction, the following system of equations is obtained

\begin{equation}
k_z^2 A + j k_z \alpha B = \varepsilon_0 \frac{\omega^2}{c^2}A \;
\label{equation_85}
\end{equation}

\begin{equation}
jk_z\alpha A -\alpha^2B = \varepsilon_0 \frac{\omega^2}{c^2}B \;
\label{equation_86}
\end{equation}

\begin{equation}
k_z^2 C - j k_z \alpha' D = \varepsilon \frac{\omega^2}{c^2}C \;
\label{equation_87}
\end{equation}

\begin{equation}
-jk_z\alpha' C -\alpha'^2D = \varepsilon \frac{\omega^2}{c^2}D \;,
\label{equation_88}
\end{equation}

and because of continuity and since a common point for both environments ({\it i.e.}, vacuum and metal) exists, it is possible to establish the conditions $A=C$ and $\varepsilon_0 B=\varepsilon D$, where $A,B,C,D$ are constants, $\varepsilon_0=1$ and $\varepsilon$ is referred to the free space and metallic environments respectively. Thus, the solution to the set of equations in (\ref{equation_85}) to (\ref{equation_88}) yields

\begin{equation}
k_z=\sqrt{\frac{\varepsilon}{1+\varepsilon}}\frac{\omega}{c} \;
\label{equation_89}
\end{equation}

\begin{equation}
\alpha=\sqrt{\frac{-1}{1+\varepsilon}}\frac{\omega}{c} \;
\label{equation_90}
\end{equation}

\begin{equation}
\alpha'=\sqrt{\frac{-\varepsilon^2}{1+\varepsilon}}\frac{\omega}{c} \;.
\label{equation_91}
\end{equation}

From (\ref{equation_89}) to (\ref{equation_91}) it can be  derived that if $\varepsilon$ is a real positive number both decay constants $\alpha$ and $\alpha'$ are imaginary numbers, and the waves do not decay with the distance in the $x$ axis. On the other hand, if $\varepsilon<1$ or is a purely imaginary number, both $\alpha$ and $\alpha'$ tends to infinite and the wave is a TM surface wave (i.e., is bounded to the surface while travelling along the $z$ axis). Thus, this TM wave can occur principally in metals or in other materials with non positive permittivity, but not in dielectric or non conductive materials.  \newline

It is well know that the permittivity of a material is given by

\begin{equation}
\varepsilon=1-\frac{j\sigma}{\omega\varepsilon_0}\;
\label{equation_92}
\end{equation}

being $\sigma$ the conductivity of the material the following

\begin{equation}
\sigma=\frac{nq^2\tau}{m(1+j\omega\tau)}\;,
\label{equation_93}
\end{equation}

where $q$ and $m$ are the electron charge and mass, respectively, and $\tau$ represents the \textit{electrons collision time}. At microwaves frequencies $w<<\tau$ so the imaginary part of $\sigma$ is negligible while its real part becomes very large. Thus, for good conductors at microwaves frequencies $\alpha$ is large, and the standing wave generated by the surface wave extends deep into the free space. From the $\alpha'$ value of the material, an indirect measurement of the surface impedance of a plane sheet can be taken.

\section{The Plasmonic Analysis}
\label{section 4}

The analysis developed in previous sections consider the presence of surface electromagnetic waves linked to the inner contour of a radially-corrugated waveguide of circular section, which is the most common setup in radioastronomy feedhorns nowadays. 

Thus, this can be analyzed by using the classical electromagnetic theory, as it has been already done, by using optics and photonics, or by using plasmonics in both optic and microwave ranges. Since it contains relevant conclusions and similarities, in this section this alternative will be briefly introduced.

Hence, in order to obtain the necessary background for the following analysis, the concept of the {\it plasmon} will be presented first.

\subsection{The Concept of Plasmon}

Plasmons are quasiparticles and the result of the quantization of the oscillations of  \textit{free electron gas} density plasma. Their properties can be derived from the Maxwell's Equations, so the relation between surface plasmons and surface waves is direct.

Surface plasmons are plasmons linked to a surface, and analogously with the surface waves they only can exist on the imaginary space between a material with real positive value of $\varepsilon>0$ (free space, air or dielectrics) and materials with $\varepsilon<0$ (conductors). 

\textit{Surface plasmon polaritons} (SPPs) are infrared or optical electromagnetic waves travelling along a metal-dielectric junction. The term \textit{surface plasmon polariton} explains how the wave involves both electrical charge motion in the metal ({\it i.e.}, surface plasmons) and electromagnetic waves in the dielectric, called \textit{polaritons}. A polariton is a quasiparticle which results from strong coupling between electromagnetic waves and electromagnetic dipole-carrying excitation. Thus, plasmons can interact with photons to create \textit{plasmon polaritons} and SPPs.

Since SPPs are defined at optical frequencies, the possibility of designing surfaces which are able to generate the counterpart of SPPs at microwaves  frequencies which are called \textit{spoof-SPPs} has been proposed (e.g. see \cite{Navarro}).

\begin{figure}[ht!]
		\includegraphics[width=0.6\textwidth]{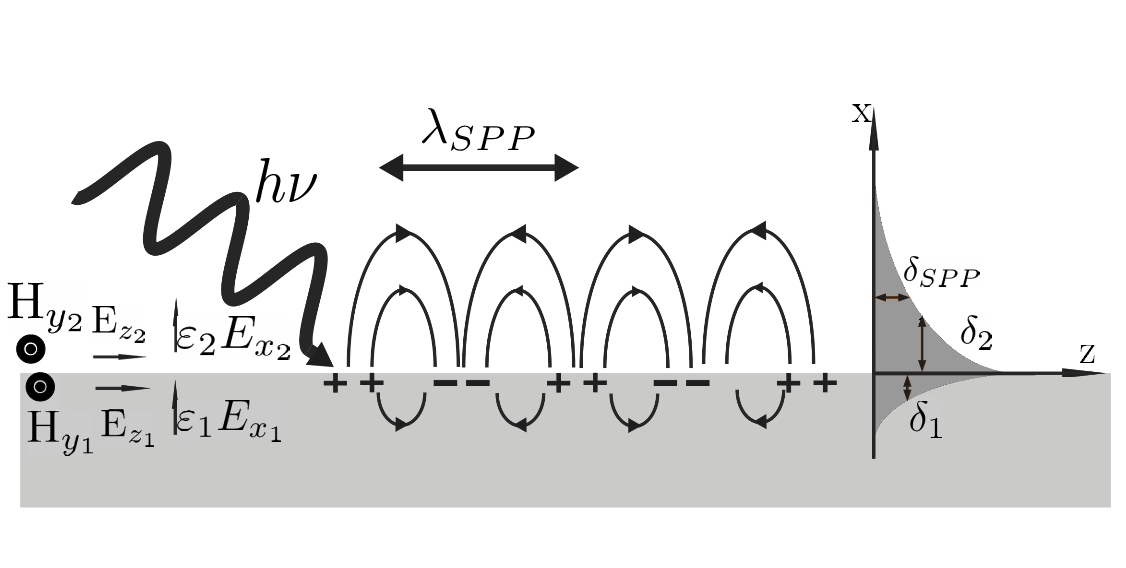}
		\centering 
		\caption{Schematic representation of an electron density wave propagating along the metal-dielectric interface.}
		\label{fig5}
	\end{figure}
	
In Fig. \ref{fig5} the charge density oscillations and associated electromagnetic fields are represented. The curves on the right represents the exponential dependence of the electromagnetic field intensity on the distance away from the interface, analogous to the case of surface waves and the decay skin depth or the attenuation constant ($\alpha$) in (\ref{equation_90}).

\subsection{The Dispersion Relation for Periodic Surfaces}

\begin{figure}[ht!]
		\includegraphics[width=0.4\textwidth]{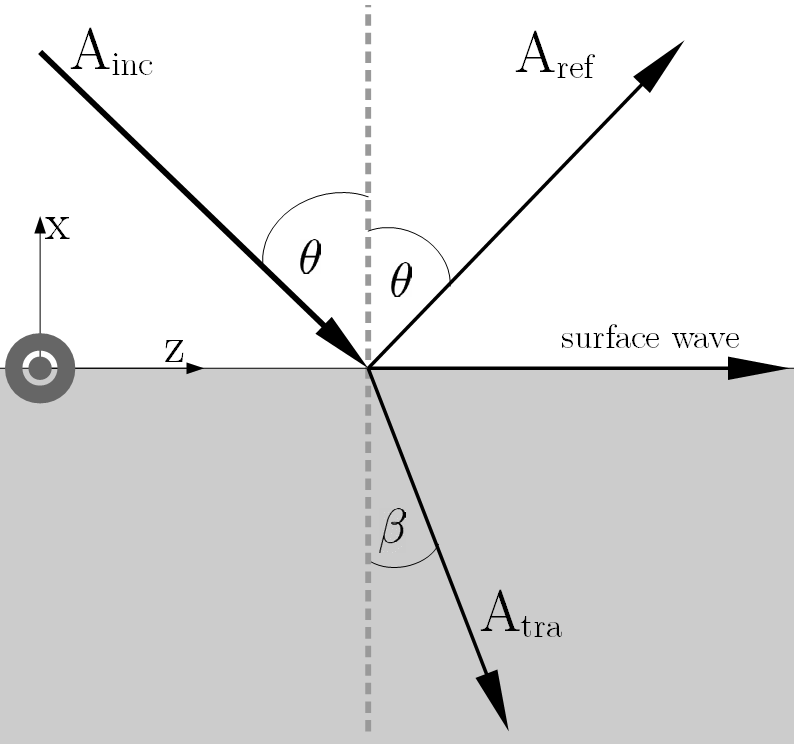}
		\centering 
		\caption{Schematic of the energy conservation of the planar obliquely incident wave in the case of coupling of surface waves.}
		\label{fig6}
	\end{figure}

The case of the reflection and transmission of plane waves obliquely incident on a planar interface surface between two media is well described in the optics bibliography by using Snells's Law\footnote{Establishes ${\frac  {\sin \theta _{1}}{\sin \theta _{2}}}={\frac  {v_{1}}{v_{2}}}={\frac  {\lambda _{1}}{\lambda _{2}}}={\frac  {n_{2}}{n_{1}}}$, where $n_i$ is the refraction index of the i-medium.} and Fresnel's Coefficients\footnote{Establishes $R_p=\left[\frac{n_1cos\theta_{1}-n_2cos \theta_{2}}{n_1cos\theta_{1}+n_2cos\theta_{2}}\right]^2$, where $R_p$ is the reflection index for the TM (also known as p-polarized) case.} for both TE and TM polarizations of the incoming radiation. In this classic analysis, an incident plane wave is totally or partially reflected or transmitted into the adjacent medium, and the coupling of surface waves is not traditionally considered. In the analysis developed in this section an incident plane wave generates three components by interaction with a  discontinuity, {\it i.e.}, a reflected wave, a transmitted wave and a surface wave linked to its surface. This situation is represented in Fig. \ref{fig6}.

In Fig. \ref{fig6} the case of incidence of a plane wave in the interface between two psuedo-infinite, isotropic and loss-less mediums is represented. Both dielectric constants $\varepsilon$ are considered complex numbers, and since both materials are non magnetic materials it is considered that $\mu=1$. The study of this situation for both SPPs and surface waves is commonly carried out using the derivation of the  \textit{dispersion relation}. This procedure is related in the following text.\newline

The field components can be expressed using vector notation 

\begin{equation}
\vec{E}=[E_x,0,E_z]e^{j(k_xx\vec{u}_x+k_zz\vec{u}_z-\omega t)}\;
\label{equation_94}
\end{equation}

\begin{equation}
\vec{H}=[0,H_y,0]e^{j(k_xx\vec{u}_x+k_zz\vec{u}_z-\omega t)}\;.
\label{equation_95}
\end{equation}

Since $D=\varepsilon E$, Maxwell's Equations can be applied to (\ref{equation_94}) and (\ref{equation_95}) obtaining 

\begin{equation}
H_y(k_z\vec{u}_z-k_x\vec{u}_x)=\varepsilon \omega (E_x\vec{u}_x+E_z\vec{u}_z)    \;,
\label{equation_96}
\end{equation}

and 

\begin{equation}
H_y=-\frac{\varepsilon \omega E_x}{k_z}= \frac{\varepsilon \omega E_z}{k_x}  \;,
\label{equation_97}
\end{equation}

and finally 

\begin{equation}
E_x=-E_z\frac{k_z}{k_x}  \;.
\label{equation_98}
\end{equation}

Here free space is named as medium $1$ and the metal as medium $2$, and the equations of the reflected, incident and transmitted fields are rewritten by using (\ref{equation_97}) and (\ref{equation_98}). 

\begin{equation}
\vec{E}_1^{inc}=E_{z_1}^{inc}\bigg[\frac{-k_{z}}{k_{x_1}},0,1\bigg]e^{j(k_{x_1}\vec{u}_x+k_z\vec{u}_z-\omega t)}  \;
\label{equation_99}
\end{equation}

\begin{equation}
\vec{E}_1^{ref}=E_{z_1}^{ref}\bigg[\frac{k_{z}}{k_{x_1}},0,1\bigg]e^{j(k_{x_1}\vec{u}_x+k_z\vec{u}_z-\omega t)}  \;
\label{equation_100}
\end{equation}

\begin{equation}
\vec{E}_2^{trans}=E_{z}^{trans}\bigg[\frac{-k_{z}}{k_{x_2}},0,1\bigg]e^{j(k_{x_2}\vec{u}_x+k_z\vec{u}_z-\omega t)}  \;
\label{equation_101}
\end{equation}

\begin{equation}
\vec{H}_1^{inc}=E_{z_1}^{inc}\bigg[0,\frac{\varepsilon \omega}{k_{x_1}},0\bigg]e^{j(k_{x_1}\vec{u}_x+k_z\vec{u}_z-\omega t)}  \;
\label{equation_102}
\end{equation}

\begin{equation}
\vec{H}_1^{ref}=E_{z_1}^{inc}\bigg[0,-\frac{\varepsilon \omega}{k_{x_1}},0\bigg]e^{j(k_{x_1}\vec{u}_x+k_z\vec{u}_z-\omega t)}  \;
\label{equation_103}
\end{equation}

\begin{equation}
\vec{H}_1^{tra}=E_{z}^{tra}\bigg[0,\frac{\varepsilon \omega}{k_{x_2}},0\bigg]e^{j(k_{x_2}\vec{u}_x+k_z\vec{u}_z-\omega t)}  \;.
\label{equation_104}
\end{equation}

In Fig. \ref{fig6} it can be seen that the surface wave (and so the SPP) travels across the interface which is common to both media, so the following relationship can be applied

\begin{equation}
E_{z_1}=E_{z_2}=E_z \, \, \, and \,\, \, H_{y_1}=H_{y_2}=H_y  \;.
\label{equation_105}
\end{equation}

The combination (\ref{equation_97}) and (\ref{equation_105}) yields 

\begin{equation}
H_y=\frac{\varepsilon_1 \omega E_{z_1}^{ref}}{k_{x_1}}= -\frac{\varepsilon_2 \omega E_{z_2}^{tra}}{k_{x_2}} \rightarrow \frac{\varepsilon_1 }{k_{x_1}}= -\frac{\varepsilon_2 }{k_{x_2}} \;.
\label{equation_106}
\end{equation}

The relationship in (\ref{equation_105}) also yields the expression of the \textit{conservation of momentum} parallel to the surface interface

\begin{equation}
k_{z_1}=k_{z_2}=k_z  \;,
\label{equation_107}
\end{equation}

and since $k_0^2=k_x^2+k_y^2+k_z^2$, this implies

\begin{equation}
k_{x_i}=\sqrt{\varepsilon_i k_0^2-k_z^2}  \;,
\label{equation_108}
\end{equation}

where the suffix $i=1,2$ denotes the medium of propagation.\newline

The combination of (\ref{equation_107}) and (\ref{equation_108}) yields

\begin{equation}
k_z=k_0 \sqrt{ \frac{\varepsilon_1 \varepsilon_2}{\varepsilon_1 + \varepsilon_2}}  \;.
\label{equation_109}
\end{equation} \newline 
The SPP dispersion relation is established from $k_z=k_{SPP}$, where it is important to note that $\varepsilon_i$\footnote{$\varepsilon_i=\varepsilon_i'+j\varepsilon_i''$ is frequency dependent. On an air-metal interface, $\varepsilon_1$ can be considered constant because its dispersion is negligible because of the lack of electronic resonances while $\varepsilon_2$ has a strong dependence on the frequency.} is a complex number, so $k$ has also both real and imaginary components. Over the optical range, the value of $\varepsilon_i$ can be approximated as a real number because of the characteristics of the materials in relation to the wavelength, but at microwaves frequencies $\varepsilon_i$ has a negligible real part and becomes a large purely imaginary positive number.\newline

Thus, for the case of a air-metal interface over the microwaves range, (\ref{equation_109}) implies that $ k_z\simeq k_0\sqrt{\varepsilon_0}<<k_0 $, so  \textit{extra} momentum is required for coupling a SPP to the surface. Different authors have studied two methods to give an excess of momentum to the wave by evanescent decay, known as \textit{prism coupling} and \textit{diffraction-grating coupling} ({\it e.g.} see \citet{Hibbins}, \citet{mlj} and \citet{Brock} dissertations). Since a corrugated horn is analogous to a grated surface, the interest here is focused on the second method.

\subsection{Diffraction-grating Coupling for Spoof-SPPs}

\begin{figure}[ht!]
		\includegraphics[width=0.5\textwidth]{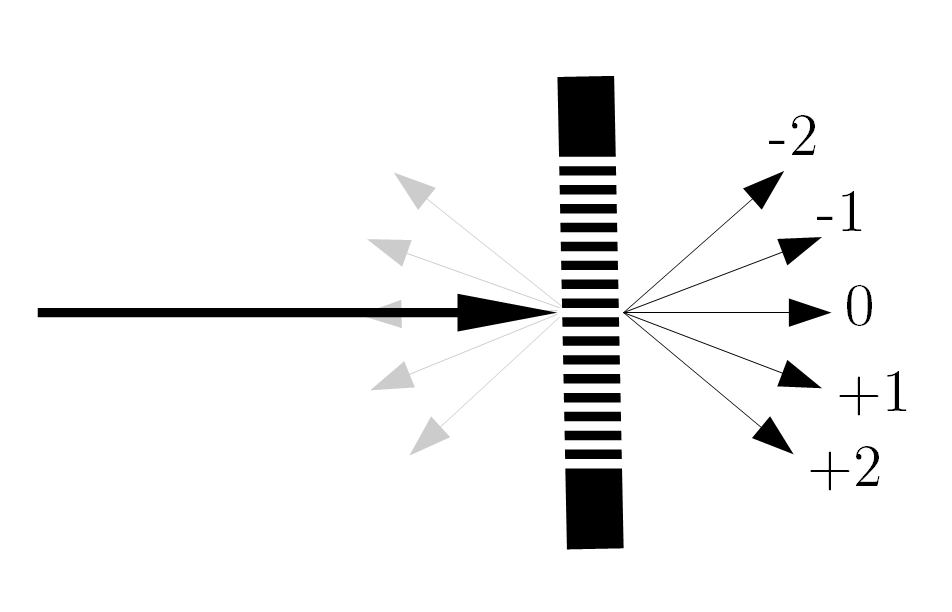}
		\centering 
		\caption{Diffraction scheme.}
		\label{fig7}
	\end{figure}

In optical theory, when a beam of light is incident upon a grating, it is considered to result in being diffracted into several different beams, which establish the orders of diffraction. This situation is represented in Fig. \ref{fig7} for 2 generic diffraction orders.

The well known {\it grating equation} can be expressed in the following form 

\begin{equation}
n_2 sin\theta_2=n_1 sin\theta_1 \pm m\frac{\lambda_0}{\lambda_p}cos\phi \;,
\label{equation_110}
\end{equation}

where $n_i$ are the refraction indices of both collindant i-media\footnote{Actually, for the junction in medium 2 an equivalent refraction index must be calculated.}, $\lambda_p$ is the physical distance between gratings (equivalent to the parameter $p$ of Fig. \ref{fig3}), $\theta_1$ and $\theta_2$ are the angles of the incident and diffracted beams with respect to the normal of the grating surface, $\phi$ is the angle of the incoming beam with the direction vector of the grating array measured in yz-plane and $m$ is the integer order of the diffraction. For example, in the case of Fig. \ref{fig7} the relative values are $\theta_1=0$, $m=0,1,2$ and $\phi=0$ since all beams are coplanar.\newline

\begin{figure}[ht!]
		\includegraphics[width=0.25\textwidth]{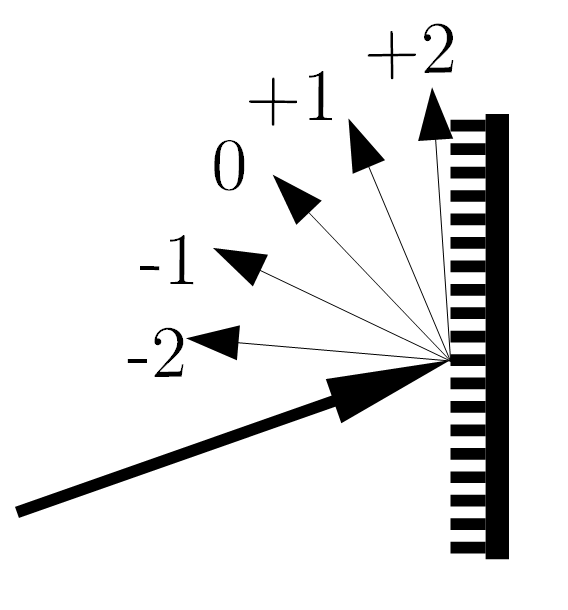}
		\centering 
		\caption{Diffraction-grating scheme.}
		\label{fig8}
	\end{figure}

On the other hand, Fig. \ref{fig8} represents the analogous situation for the case of a reflective diffraction grating, for example a corrugated metal. In accordance with the analysis previously taken for Fig. \ref{fig6}, part of the radiation is transmitted into the surrounding media. The equation which describes the situation of Fig. \ref{fig8} is (\ref{equation_110}).\newline

In Fig. \ref{fig8}, the beam of order $m=2$ is close to being diffracted in the parallel direction to the grating surface. This situation in the optical range could be \textit{easily} achieved by adjusting the angle $\theta_1$ and the distance $\lambda_p$ for a given wavelength $\lambda_0$. 

In addition to this, to couple an SPP a specific value of the wavevector is needed. The value of this wavevector can be derived by transforming (\ref{equation_110}) converting the grating formula to wave vectors, resulting in (\ref{equation_113}). In order to achieve this conversion, firstly it is necessary to multiply all the terms in the equation by $2\pi/\lambda_0$, and then to apply the definition of the wavevector. This leads to

\begin{equation}
\frac{2\pi}{\lambda_0}n_2 sin\theta_2=n_1 k_0 sin\theta_1 \pm m\frac{2\pi}{\lambda_p}cos\phi \;.
\label{equation_111}
\end{equation}

\begin{figure}[ht!]
		\includegraphics[width=0.8\textwidth]{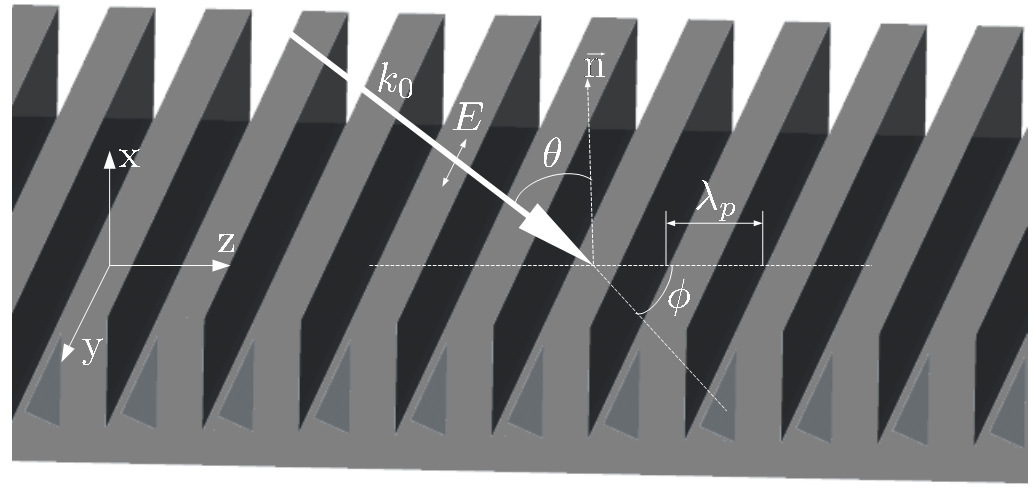}
		\centering 
		\caption{Diffraction-grating coupling scheme.}
		\label{fig9}
	\end{figure}

Since it is possible to define a wavevector $k_p=2\pi /\lambda_p$ and for the coplanar case $cos\phi=1$, it is possible to identify common terms between (\ref{equation_111}) and  (\ref{equation_113}). However, a more general case is shown in Fig. \ref{fig9}. \newline

Figure \ref{fig9} represents the case of TM polarized radiation\footnote{Also called p-polarized as opposed to  s-polarized, which is equivalent to a TE polarization.} hitting a semi-infinite grating with an incident angle $\theta$ with the $yz$-plane and an angle $\phi$ with the $xz$-plane.

By introducing corrugations on the schematic of Fig. \ref{fig6} to arrive to the case of Fig. \ref{fig9} case, the translational symmetry of the surface vanishes and the tangential momentum of an incident photon is no more conserved. The periodicity given by $\lambda_p$ produces diffraction orders. When an order gives rise to a wavevector $k_p$ larger than the maximum permissible momentum for the incident radiation in medium $1$ it becomes evanescent. The wavevector of evanescent fields related with non propagated diffraction orders makes incident radiation coupling SPPs only if the following condition is satisfied

\begin{equation}
k_{SPP}^2=n_1^2 k_0^2\,sin^2\theta + m^2 k^2_p\pm 2 {n_1} m\,k_p\,k_0\, sin\theta \, cos\phi \;,
\label{equation_112}
\end{equation}

where $m$ is the order, $\sqrt{\varepsilon_{r_1}}=n_1$ corresponding to the relative refractive index of medium $1$ and $k_p=2\pi/\lambda_p$. \newline

For the particular case of null azimuth angle ($\phi=0$), hence $k_{SPP}$, $k_0$ and $k_p$ are collinear and (\ref{equation_112}) is simplified to

\begin{equation}
k_{SPP}=n_1k_0\,sin\theta \pm m k_p  \;.
\label{equation_113}
\end{equation}

For the situation represented in (\ref{equation_113}) it is established that $\phi=0$ and only TM polarised radiation can excite SPPs. On the other hand, for the case in (\ref{equation_112}), with values $0<\phi < \pi/2$, both TM and TE polarized radiation produce SPPs. This is analyzed in the following paragraph.\newline

\begin{figure}[ht!]
		\includegraphics[width=0.9\textwidth]{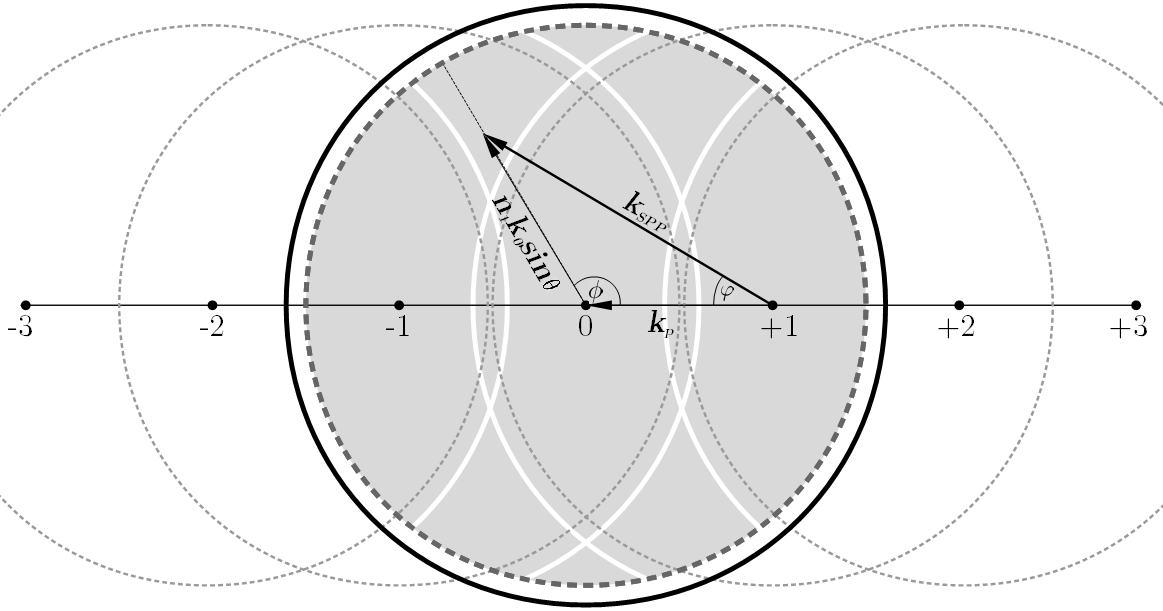}
		\centering 
		\caption{Coupling SPPs modes scheme.}
		\label{fig10}
	\end{figure}
	
 Fig. \ref{fig10} represents a \textit{k-space}\footnote{Relation of particle energy to momentum.} generic representation of the possible solutions for coupling SPPs given from (\ref{equation_112}). The gray dashed-radius represents the maximum possible in-plane momentum for the propagation of a photon above the grating. The circle  with a black solid radius line represents the SPP wavevector ($k_0$) of the 0-th order SPP-mode, which is larger than the incident wavevector ($k_0sin\theta$) required for the excitation of a SPP in a $k_p$ scattering process, while the white arcs which cross the shaded area represent the i-th diffraction orders for $i>0$. Coupling SPPs is only possible inside the shaded area, whose limit is the traced line. The limits of the shaded areas for the i-th diffraction orders are also represented. The vectors represent the incidence of a photon with angle $\theta$ with respect to the normal vector $\vec{n}$ and an azimuth angle $\phi$ coupling an SPP which is propagated with an angle $\varphi$ with respect to the $\vec{u_z}$ direction via a grating vector $k_p$. The grating allows the SPP-wavevector ($k_{SPP}$) to be changed by integer $nk_p$ multiples (only the $n=1$ case is represented in the Fig. \ref{fig10}).\newline

It is possible to extend the analysis of the SPP coupling for the case of 2-dimensional gratings, where an array of grooves with lattice $\lambda_p$ exists through the $\vec{u}_z$ direction and an array of lattice $\lambda_g$ exists on the $\vec{u}_y$ direction. This is

\begin{equation}
\vec{k}_{SPP}=\delta\sqrt{\varepsilon_1}k_0\,sin\theta  \vec{u}_{yz} \pm m_z k_p \vec{u}_z  \pm m_y k_g\vec{u}_y \;,
\label{equation_114}
\end{equation}

where $\delta=1$ for the case of TM polarization and $\delta=0$ for TE polarization.

\section{Soft and Hard Surfaces}
\label{section 5}

Apart from the analyses presented before this section, \citet{Kildal} carried out some research defining two types of electromagnetic surfaces, i.e., \textit{soft} and \textit{hard} boundaries. Since both analyses are convergent and lead to analogous conclusions, it is only necessary to establish the connection between corrugated feed horns and soft surfaces. This will serve as a basis for concepts which will be used in later sections.

Kindal defines a soft surface as

\begin{equation}
SOFT
\begin{cases} 
 |X^{TM}\big|\rightarrow \infty \; \Leftrightarrow\; H^{TE}\rightarrow 0 \\
 |X^{TE}\big|\rightarrow 0  \; \Leftrightarrow\; E^{TE}\rightarrow 0 \;,
\end{cases} 
\label{equation_115}
\end{equation}

And a hard surface  as

\begin{equation}
HARD
\begin{cases} 
 |X^{TM}\big|\rightarrow  0 \;\Leftrightarrow\; H^{TE}\rightarrow \infty \\
 |X^{TE}\big|\rightarrow 0  \; \Leftrightarrow\; E^{TE}\rightarrow 0 \;.
\end{cases} 
\label{equation_116}
\end{equation}

where $X^i$ is the surface reactance of the $i$-polarization. Here it is again assumed that for metallic surfaces $Z^i\simeq X^i$ since $R^i \simeq 0$.\newline

Ideal soft and hard surfaces do not exist practically, but they can be approximated by surfaces with $\big|X^{TM}\big|>>\big|X^{TE}\big|$ for the case of soft boundaries and with $\big|X^{TM}\big|<<\big|X^{TE}\big|$ for the hard boundaries case.

\subsection{Planar Corrugated Soft and Hard Surfaces}

Consider a radially corrugated surface such as that represented in Figs. \ref{fig3} and \ref{fig9}, which satisfies the condition

\begin{equation}
t<<p<<\frac{\lambda_0 }{2}\sqrt{\varepsilon_{b}}\;,
\label{equation_117}
\end{equation}

where $\sqrt{\varepsilon_{b}}=n_{b}$ is the relative refractive index of the material which fills the space inside the grooves (not necessarily air since it is possible to fill this space with a different dielectric), and where again the parameter $t$ is the width of the corrugation, $d$ its depth and $b+t=p$ is the lattice period, and, as is represented in the Fig. \ref{fig9}, a surface wave is traveling along the $z$ axis. Thus, the $E_z$ field satisfies

\begin{equation}
E_z=sin(k_0x+k_0d)\;,
\label{equation_118}
\end{equation}

where $k_0=2\pi/\lambda_0$. By applying the Maxwell-Faraday Equation\footnote{In its form $\vec{\nabla}\times\vec{E}=j\omega\mu\vec{H}$.} at the plane of the top of the groove ($x=0$), the soft condition (\ref{equation_115}) becomes equivalent to

\begin{equation}
\frac{\partial E_z}{\partial x}=0\;,
\label{equation_119}
\end{equation}

which is satisfied for the case of $d=\lambda_0/4$. Hence, a corrugated planar structure with the restrictions in (\ref{equation_117}) is a soft surface if $d=\lambda_0/4$. This condition has been already established from the \textit{classical} derivation in section \ref{section 3}. \newline 

\begin{figure}[ht!]
		\includegraphics[width=0.4\textwidth]{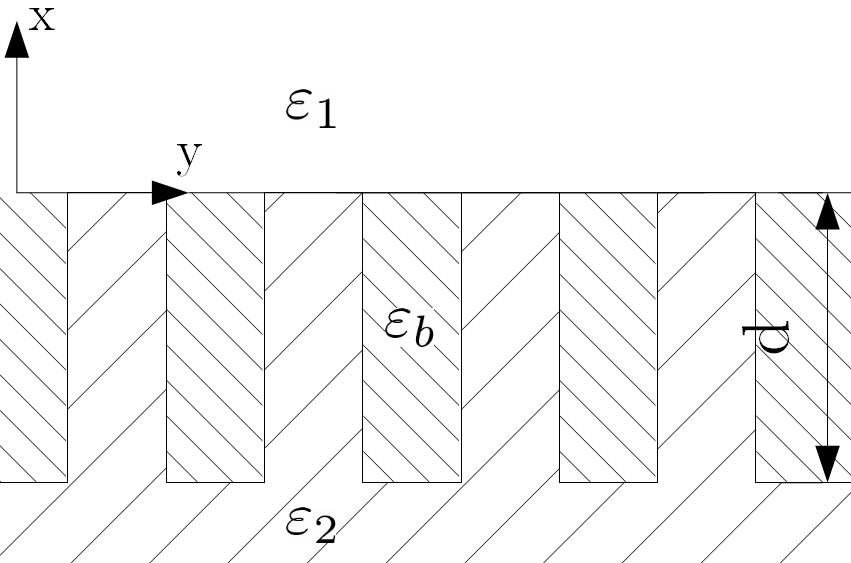}
		\centering 
		\caption{Longitudinally corrugated surface.}
		\label{fig11}
	\end{figure}

On the other hand, for the case of a longitudinally corrugated surface (with the corrugation rotated $90^{\circ}$ so $x\leftarrow x$, $-y\leftarrow z$ and $z\leftarrow y$ once the rotation is completed) as is shown in Fig. \ref{fig11}, the $E_y$ field satisfies

\begin{equation}
E_y=sin(k_xx+k_xd)e^{-jk_zz}\;,
\label{equation_120}
\end{equation}

where\footnote{See (\ref{equation_108}).}

\begin{equation}
k_x=\sqrt{\varepsilon_bk_0^2-k_z^2}=k_0\sqrt{\varepsilon_b-sin^2\theta}\;,
\label{equation_121}
\end{equation}

and

\begin{equation}
k_z=k_0sin\theta=k_0(\vec{u}_{x}\cdot \vec{u}_z)\;,
\label{equation_122}
\end{equation}

$\theta$ being the angle of incidence.

In (\ref{equation_120}), $E_y=0$ when $x=-d$, and $H_z=0$ at the top ($x=0$). Analogously to the previous case for obtaining (\ref{equation_119}), from Maxwell`s Equations the following condition is obtained

\begin{equation}
\frac{\partial E_y}{\partial x}=0\;,
\label{equation_123}
\end{equation}

at the top surface ($x=0$), from (\ref{equation_116}) it can be seen that the surface is hard when $k_x d =\pi/2$, or equivalently

\begin{equation}
d=\frac{\lambda_0}{4\sqrt{\varepsilon_b-sin^2\theta}} \;.
\label{equation_124}
\end{equation}

From (\ref{equation_124}), a longitudinally corrugated surface can't be a hard-boundary for $\theta\neq \pi/2$ if $\varepsilon_b\ne1$. That is, for angles of incidence not parallel to the surface, it is necessary to fill the spaces between grooves with a material with an relative dielectric constant larger than the air ($\varepsilon_b>>1$) in order to obtain a hard surface.

\subsection{The Soft and Hard Analysis Applied to Cylindrical Corrugated Surfaces}

Here, the case of a corrugated cylindrical waveguide will be analyzed. Fig. \ref{fig2} represents the highlighted box of the corrugated feedhorn in Fig. \ref{fig3}. This structure is equivalent to a cylindrical transversely corrugated waveguide, which can be treated as quasi-infinite.\newline

For this waveguide, the electric field is given by the expression

\begin{equation}
E_z=cos\phi[J_1(kr)+CN_1(kr)] \;,
\label{equation_125}
\end{equation}

where $J_1$ represents the Bessel function of the first kind and of first order, $N_1$ is the Newman function of first order and $C$ is a constant determined by the condition $E_z=0$ at the surface $r=a$. Thus

\begin{equation}
C=\frac{-J_1(ka-kd)}{N_1(ka-kd)} \;.
\label{equation_126}
\end{equation}

This boundary is soft when $d$ has a value which makes $H_y\equiv H_{\phi}=0$ at the surface given by $r=a$, which defines the aperture radius. Hence, the boundary is soft when the following equation is satisfied

\begin{equation}
J'_1(ka)N_1(ka+kd)-N'_1(ka)J_1(ka+kd)=0 \;.
\label{equation_127}
\end{equation}\newline

The solutions of the condition represented in (\ref{equation_127}) indicate that $d\rightarrow\lambda/4$ when $a$ increases. From (\ref{equation_125}) it can be derived that $H_z$ depends on $k$, and $k^2-k_z^2=k_r^2$, $k_r$ being inversely proportional to $r_0$, and $k^2$ proportional to $1/a$. Thus, by using again the Maxwell-Faraday Equation in its cylindrical form we find

\begin{equation}
E_z=j\frac{\eta}{k}\frac{1}{a}\frac{\partial H_r}{\partial \phi}  \rightarrow E_r\propto \frac{1}{ka^3}\;.
\label{equation_128}
\end{equation}

The relation (\ref{equation_128}) can be applied for both transverse fields ({\it i.e.}, $E_r\equiv E_x$ or $E_r \equiv E_y$). Therefore, the conditions for supporting the hybrid-mode in a cylindrical horn are that $d\rightarrow\lambda/4$ and $ka\rightarrow \infty$, as has been already established in the section \ref{section 3} by a \textit{traditional} methodology.\newline

\begin{figure}[ht!]
		\includegraphics[width=0.55\textwidth]{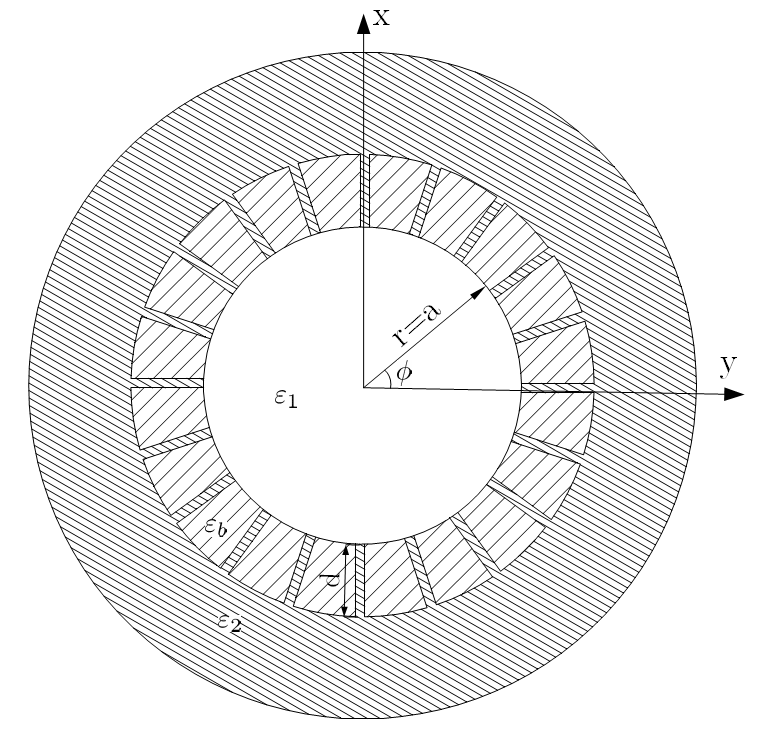}
		\centering 
		\caption{Longitudinal cylindrical waveguide xy-cut.}
		\label{fig12}
	\end{figure}

Analogously, Fig. \ref{fig12} represents a transverse cut for the case of a longitudinally corrugated cylindrical waveguide, in order to analyze the characteristics of a feedhorn with a hard surface boundary. Here it is supposed that the grooves are filled with a dielectric material different from that of air.

The magnetic field satisfies 

\begin{equation}
H_z=[J_0(k_rr)+CN_0(k_rr)]e^{-jkz}\;,
\label{equation_129}
\end{equation}

where $k_r=k\sqrt{\varepsilon_b-1}$, $J_0$ and $N_0$ are the Bessel and Newman functions of order zero and $C$ can be obtained from the following condition

\begin{equation}
E_{\phi} =0 \leftrightarrow \frac{\partial H_z}{\partial r} =0 \;.
\label{equation_130}
\end{equation}

Thus, the requirement in order to make this surface a hard boundary is given by

\begin{equation}
J_0(k_ra)N_1(k_ra+k_rd)-N_0(k_ra)J_1(k_ra+k_rd)=0\;.
\label{equation_131}
\end{equation}

By employing Maxwell`s Equations in cylindrical coordinates for the case of a null charge density and for the modes $TM_{11}$, $HE_{11}$ and $TE_{11}$, (modes with $E_r\propto sin\phi$ and $E_{\phi}\propto cos\phi$),  the boundary conditions of $E_r$ and $E_{\phi}$ satisfy\newline

\begin{equation}
\frac{\partial (rE_r)}{\partial r}=-\frac{\partial E_{\phi}}{\partial\phi} \;
\label{equation_132}
\end{equation}

\begin{equation}
\frac{\partial (rE_{\phi})}{\partial r}=\frac{\partial E_r}{\partial\phi} \;.
\label{equation_133}
\end{equation}
\newline The equations (\ref{equation_132}) and (\ref{equation_133}) are satisfied when $E_y=E_r sin\phi+E_{\phi}cos\phi =K$ or equivalently when $E_r=K sin\phi$ and $E_{\phi}=K cos\phi$, $K$ being an arbitrary constant.

This implies that a TEM mode can exist inside the longitudinally corrugated waveguide satisfying $E_y=E_r sin\phi+E_{\phi}cos\phi =K$ and $E_x=0$.

Since $k_r=k\sqrt{\varepsilon_b-1}$ it is neccessary for  $\varepsilon_b>1$ to support this TEM mode, which only exists when $d$ is a suitable value (in wavelengths). Since $TE_{11}$, $TM_{11}$ and $HE_{11}$ modes degenerate to TEM when $k_z\rightarrow k$ and $k_r\rightarrow 0$, the TEM mode inside the cylinder can be seen as a degenerate mode. Outside of the operating band where $d$ is not a suitable value, a degenerated $TE_{11}$ exists. 

\begin{figure}[ht!]
		\includegraphics[width=0.5\textwidth]{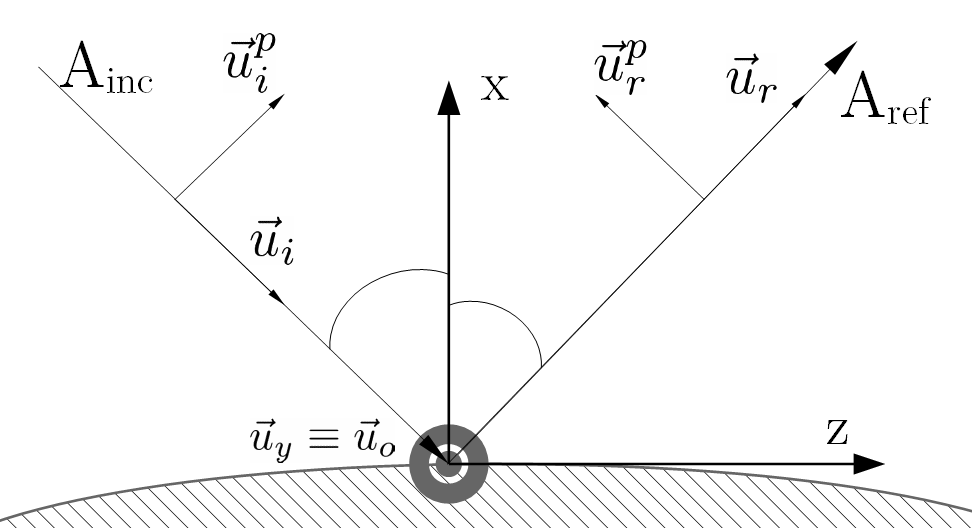}
		\centering 
		\caption{Geometric Optics reflections analysis.}
		\label{fig13}
	\end{figure}

Thus, it is theoretically possible to design a cross section of a longitudinally corrugated horn with null cross-polarization. The field distribution on this kind of waveguide will have a constant value and straight field lines corresponding to zero cross-polarization.\newline
\newline In conclusion, it is possible to design horns with zero cross-polarization by employing both soft and hard surfaces, since they do not create cross polarization due to geometric optic (GO) reflections because the reflection coefficients of hard and soft surfaces are equal to $\pm1$, respectively, for both directions of the transverse field. This GO interpretation will be presented in the following paragraph.\newline
\newline Let's consider the incident and reflected rays in Fig. \ref{fig13}, where the normal direction with respect to the surface is $\vec{u_x}$, the longitudinal direction is given by $\vec{u_z}$ and the transverse-orthogonal direction is given by the vector $\vec{u_y}$, which is off-plane. 

Thus, the incident and reflected fields can be decomposed on their parallel polarization (suffix $p$) and its orthogonal polarization (suffix $o$) components. This is

\begin{equation}
\vec{E_i}=E_p \vec{u}_i^p+E_o \vec{u}_o\;
\label{equation_134}
\end{equation}

\begin{equation}
\vec{E_r}=R_p E_p \vec{u}_r^p+R_o E_o \vec{u}_o\;,
\label{equation_135}
\end{equation}

where $R_p$ and $R_o$ are the parallel and orthogonal directional reflection coefficients, respectively.\newline

By superposition of the incident and reflected rays it is possible to deduce that the boundary condition $E_y=E_z=0$, or equivalently  $Z_y=Z_z=0$, is only satisfied if $R_p=1$ and $R_o=-1$.\newline

$R_o=-1$ corresponds to a short circuit reflection, and hence with voltage $V=0$. This, according to (\ref{equation_115}) corresponds to a soft boundary. Thus, it is theoretically possible to obtain a soft surface in the case of having orthogonal and parallel (in plane) reflection coefficients with the value $R_p=R_o=-1$ independent of the polarization of the incoming radiation. In this case, it is observed by superposition in (\ref{equation_134}) and (\ref{equation_135}) that $E_y=E_x=0$ for both transverse fields, and that $E_z\neq0$.\newline

Similarly, hard surfaces independent of the polarization of the incident ray can be obtained for the case $R_p=R_o=1$. For superposition of the incident and reflected beams expressed in (\ref{equation_134}) and (\ref{equation_135}), here it is possible to deduce that this is equivalent to the condition $E_z=H_z=0$, or equivalently $|Z_y\big|\rightarrow \infty$ and $Z_z=0$. This corresponds to 

\begin{equation}
\frac{\partial E_x}{\partial x}\simeq 0\;\; and \;\; \frac{\partial E_y}{\partial x}\simeq 0 \;,
\label{equation_136}
\end{equation}

if the radii of curvature of the surface in Fig. \ref{fig13} is sufficiently large (or the surface is quasi-planar).

\section{The Rise of Metamaterials}
\label{section 6}
A commonly accepted and precise definition of {\it metamaterial} does not exist. However, metamaterials can be understood as materials with properties given by their geometrical structure which differs from that given by their basic material properties. Their geometric structures have lattice periods much shorter than the operating wavelength. \newline

According to this brief definition, the corrugated wall of a planar surface, a waveguide or a feedhorn studied in previous sections can be understood as an electromagnetic metamaterial. Throughout this section, the study of the use of different kinds of metamaterials in order to fabricate feedhorns for radioastronomy purposes will be presented, starting with the the \textit{high impedance surfaces} (HIS) based on the research of \citet{Sievenpiper2} in electromagnetic band-gap structures (EBG), along with its possible application to designing horn antennas for radioastronomy and satellite communications.\newline

The effect of incoming radiation hitting the metallic surface of the horn has already been presented in earlier sections and it has been shown that it can become polarized due to the interaction with this discontinuity. The Kildal analysis provides tools for designing materials which do not create this cross-polarized beams since their reflection coefficients are $\pm1$. Thus, by designing adequate materials (Electromagnetic meta-materials) the incoming electromagnetic radiation can preserve its original polarization passing through a guiding structure, and thus this crucial scientific information can be maintained from the source.

\subsection{The Classic Electromagnetic Metamaterials}

\begin{figure}[ht!]
		\includegraphics[width=1\textwidth]{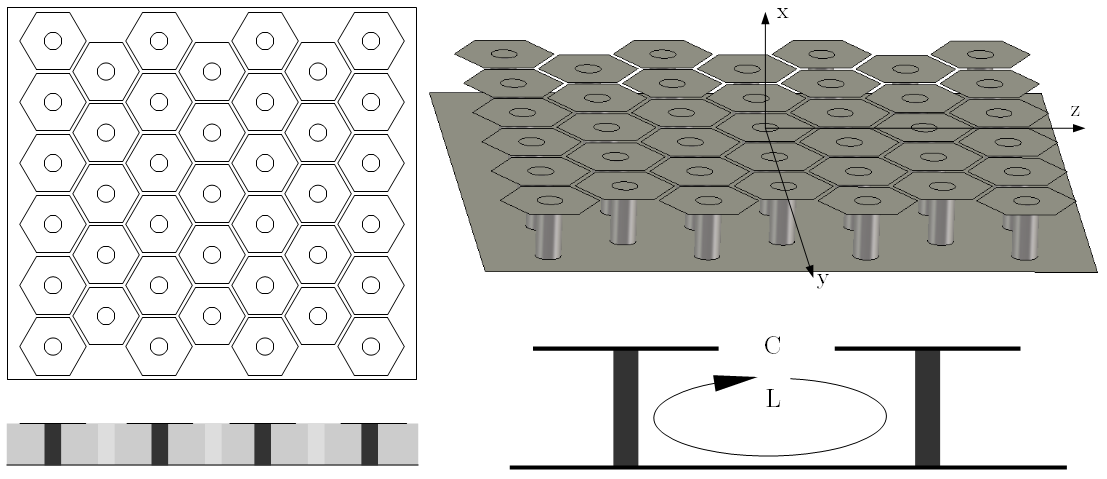}
		\centering 
		\caption{Metasurface classical mushroom pattern. Top (upper-left), front (lower-left), 3D with dielectric removed (top-right) and equivalent LC circuit (bottom-right) views.}
		\label{fig14}
	\end{figure}

 Fig. \ref{fig14} shows a model of the classic Sievenpiper mushroom structure, constituted by periodically spaced hexagonal top plates with a short circuit at the center, made by a cylindrical pin connected to the ground plate. 

The space between the top and the ground plates is filled with a dielectric material, so the fabrication by using PCB technology is feasible. By applying the \textit{effective surface impedance approximation}-model, similar to the case of the corrugated waveguide studied in the section \ref{section 3}, it is possible to associate a surface impedance value to the top plane (surface at $x=0$). \newline

Analogously to the case of the corrugated surfaces studied in the section \ref{section 3}, an equivalent parallel LC circuit model can be established (figure \ref{fig14} bottom-right). The surface impedance of this metamaterial is given by (\ref{equation_66}), and a resonant frequency exists, which is given by (\ref{equation_64}). Thus, the metasurface's impedance can be inductive, capacitive or both. 

Rememebering that a purely inductive surface produces a $+\pi/2$ reflection phase-shift and a capacitive surface a $-\pi/2$ reflection phase-shift, it is possible to establish a bandwidth in which the reflection phase-shift varies from $+\pi/2$ to $-\pi/2$ passing through $0$ rad phase-shift, which corresponds with the resonance frequency. This relative bandwidth is given by

\begin{equation}
\frac{\Delta \omega}{\omega_0}=\frac{Z_{\omega_0}}{Z_0} \;,
\label{equation_137}
\end{equation}

where $Z_{\omega_0}=\sqrt{L/C}$ is the characteristic impedance of the LC circuit and $Z_0$ is the impedance of free space. 

Within this bandwidth, a tangential transmission forbidden bandgap for TM and TE waves is found. This forbidden bandgap is centered at the resonant frequency.\newline

\begin{figure}[ht!]
		\includegraphics[width=0.8\textwidth]{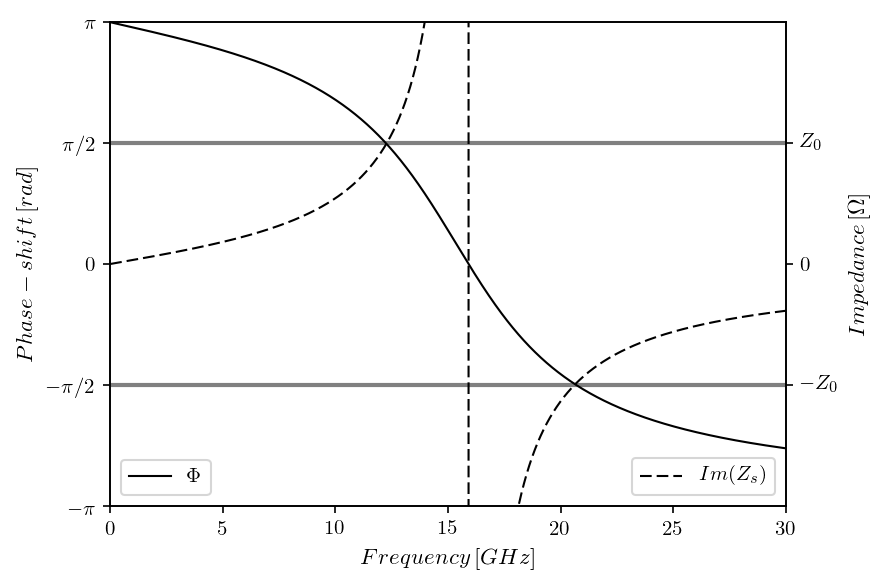}
		\centering 
		\caption{Sievenpiper mushroom theoretical behaviour over frequency with $L=2$ $nH^2$ and $C=0.05$ $pF^2$}
		\label{fig15}
	\end{figure}

Fig. \ref{fig15} represents the theoretical reflection phase-shift for a classic Sievenpiper mushroom structure, which is given by (\ref{equation_74}). Note that this pattern is approximately geometrically and electromagnetically equivalent in both $\vec{u_y}$ and $\vec{u_z}$ directions, so $L$ and $C$ are direction independent\footnote{And are commonly expressed in squared units.}. 

The bandwidth $\Delta \omega$ lies in the range between the $\pm \pi/2$ phase-shifts, and the resonant frequency is at the center of the bandwidth, around $15$ GHz in this case.\newline

Since the relative bandwidth given by (\ref{equation_137}) is proportional to $\sqrt{L/C}$, if the capacitance of the metamaterial is increased, the bandwidth decreases. 

An important point is that while the reflection phase of a metallic plane sheet is well known as destructive, with a $\pi$ rad phase-shift, it is possible to design a metasurface with a constructive reflection phase-shift of $0$ rad. This has been recently used for the designing of antennas for applications in telecommunications ({\it e.g.} see \citet{Sievenpiper2}) where surface currents in the ground plane are undesirable, causing unwanted reflections from the ground-plane edge.

\subsection{State of the Art of Horn Antennas based on Metamaterials} 

\label{Section_6.2}

Different patterns of metamaterials have been published in recent years ({\it e.g.} \citet{Yang}, \citet{Sievenpiper3}, \citet{Silveirinha}), principally for planar antenna applications in telecommunications. These works have a different goal from that of this article. The metamaterial design can be applied to improving dipole antenna grounds in Radio-astronomy but not directly to widebanding corrugated feedhorns. An example is given in Fig. \ref{fig15}, where a mushroom type metamaterial has geometrical dimensions adapted for high impedance devices and $L=2$ $nH^2$,  $C=0.05$ $pF^2$.  \newline
\newline However, several authors ({\it e.g.} \citet{Lier}, \citet{Scarborough}, \citet{Scarborough2},  \citet{Shahcheraghi} and \citet{Werner}) have recently presented novel metahorns\footnote{Understood here as a feedhorn designed to support the hybrid-mode  based on metamaterials instead of corrugations.} principally designed for satellite communications applications. Although satellite communications antennas have not always had the same design specifications as typical radioastronomy antennas referring to properties such as the optimal transverse section of a feedhorn\footnote{Satellite communications feedhorns can present rectangular sections in order to create an asymmetric pattern on transmit or match an asymmetric focal field in receive mode.} and levels of cross-polarization in the antenna pattern (see \citet{MFI}), there is the possibility
of a common design that would lead commercial applications in both fields.\newline
\newline In  section \ref{section 5} it has been established that corrugated feedhorns are soft surfaces, and that even a hard surface can lead to null cross-polarized patterns. 

The metamaterials of interest to design metahorns can be also classified as soft or hard in the Kildal sense, and for the case of a feed-horn with a low cross-polarization pattern the condition in (\ref{equation_9}) (or equivalently in (\ref{equation_49})) must be satisfied, as has been already properly established before.\newline

Thus, from (\ref{equation_9}) and since in the absence of losses\footnote{$R\simeq 0$ through a perfect metallic conductor.} a surface impedance can be expressed as a combination of an inductive reactance and a capacitive reactance by using an equivalent LC circuit and it is possible to obtain an equation which relates the values of inductance ($L$) and capacitance ($C$) for a generic metamaterial. Notice that for a spatially anisotropic metamaterial the values of $L$ and $C$ can vary with dependence on the spatial direction, so $X^{TE}=j\omega L^{TE}-j/\omega C^{TE}$ and $X^{TM}=j\omega L^{TM}-j/\omega C^{TM}$ and by substituting in (\ref{equation_49}) the hybrid-mode condition for a waveguide with anisotropic metasurface walls become

\begin{equation}
\omega^2L^{TE}L^{TM}-\frac{L^{TE}}{C^{TM}}-\frac{L^{TM}}{C^{TE}}+\frac{1}{\omega^2C^{TE}C^{TM}}=Z_0^2 \;.
\label{equation_138}
\end{equation}

For example, by substituting the values of the Sievenpiper mushroom shown in the Fig. \ref{fig14} ($L^{TE}=L^{TM}=2$ nH and $C^{TE}=C^{TM}=0.05$ pF) in (\ref{equation_138}), it is found that the hybrid-mode condition is not reached by several orders of magnitude. Hence, this metamaterial is not a good candidate for metahorns (because $L/[nH^2]>>C/[pF^2]$ and $C/[pF^2]<<freq/[Hz]$)\footnote{Actually, the Sievenpiper mushroom is a hard surface, while a corrugated waveguide is a soft surface.}. 

Thus, in order to design a metasurface which supports the field distribution reflected from a paraboloid (see section \ref{section 2}) it is necessary to force a more balanced $L/C$ ratio in the case of {\it isotropic} surfaces\footnote{Understanding as isotropic a metasurface which has the same geometry in both TE and TM spatial field directions.} or with $L^{TE}\neq L^{TM}$ and $C^{TE}\neq C^{TM}$ with adequate values for the case of anisotropic metamaterials. 

\begin{figure}[ht!]
		\includegraphics[width=0.6\textwidth]{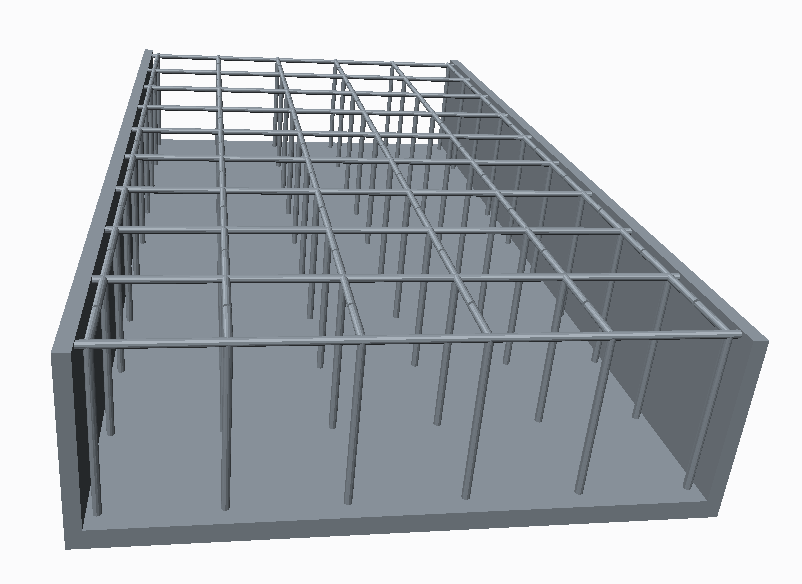}
		\centering 
		\caption{Example of metamaterial based on \citet{Lier}. Note the gaps on the wire-grid in the longitudinal direction providing $L^{TE}\neq L^{TM}$ and $C^{TE}\neq C^{TM}$ anisotropy type.}
		\label{fig16}
	\end{figure}

It is important to remember now that the condition in (\ref{equation_49}) could be satisfied by imposing a $X^{TE}\sim X^{TM}$ reactances for the case of hard surfaces such as the Sievenpiper structure or a $X^{TE} << X^{TM}$ behavior for the case of soft boundaries such as a corrugated feedhorn.\newline
\newline An illustrative example of a possible soft metamaterial based on the work by Scarborough et al. is shown in the Fig. \ref{fig16}. This metamaterial has been manufactured, inserted on a rectangular section for satellite communications horn and tested, resulting in a low cross-polarization pattern throughout the C-band.

On the other hand, the simulations\footnote{The fabrication and laboratory test of this metahorn is a technological challenge and it has not yet been completed.} in \citet{Scarborough2} leads us to think that by substituting corrugations by a novel soft modified mushroom based metamaterial in a conical feedhorn the bandwidth with low cross-polarization and sidelobe levels can be significantly increased, covering a wide 10-20 GHz band with a cross-polarization attenuation level around -35 dB. However, ring-loaded corrugated feedhorns can achieve a 2.4:1 bandwidth  too (e.g. see \citet{Clarricoats}).

There are significant advantages in weight and complexity of this wideband meta-horn for novel astronomical instruments which are under development today, where often the band must be split into $n$ sub-bands which are covered with $n$ corrugated feedhorns because of the bandwidth limitations that the corrugated horns present.

The technological goal for this article is to design a single ultra-wideband metahorn to replace several corrugated feedhorns, which in turn could potentially increase the sensitivity of a given  radiometer by $\sqrt{n}$ factor\footnote{From ideal {\it Radiometer Equation} the sensitivity of radiometers is given by $\delta T=\frac{k T_{sys}}{\sqrt{\Delta \nu t}}$, where $k$ is a factor of the order of  unity, dependent on the configuration, $T_{sys}$ is a well known figure of merit in radioastronomy, $\Delta \nu$ is the radiometer bandwidth and $t$ is the integration time.}, because feedhorns are the main limiting component in terms of bandwidth. There are wide band designs for the  other electric and optical components in a radiometer.
	
Several authors have published theoretical models for the analytical electromagnetic calculation of metamaterials structures ({\it e.g.}, see \citet{Luukkonen}). However, when a geometric change in the metamaterial is carried out, the analytical model must be revised because the analysis cannot be applied to the new shape, consuming resources and time. 

On the other hand, {\it finite difference method} calculations allow unlimited changes to the geometry of the metamaterial, with the down-side of much slower computation time. There are several specialized software packages for  electromagnetic calculation of 3D-models by using the finite differences method. Since here we desire to explore many different metamaterial geometries in search of the ideal one in terms of bandwidth, cross-polarization and the sidelobes level throughout this band, the finite difference method software simulation is the best choice.\newline

Since the condition in (\ref{equation_9}) or (\ref{equation_49}) can be achieved by maintaining $|X^{TE}|\sim |X^{TM}|\sim Z_0$ through a given band or by maintaining  $X^{TE} >> X^{TM}$ or $X^{TE} << X^{TM}$ with adequate values in this band\footnote{This is the case of soft and hard boundaries.}, a great variety of geometries can be explored, including isotropic and anisotropic cases, soft and hard surfaces, HIS, modified corrugations and others.\newline
\newline This is novel research, and there is still a large improvement to be made over already published work, so the goal now is to design a new metamaterial which provides quasi-zero cross-polarization and a low sidelobe level throughout a wider bandwidth than the classic corrugated horns achieve. The work is not trivial. There is a technological difficulty in manufacturing shaped geometries on the inside of a feedhorn which often leads to state-of-the-art mechanical techniques and then there is the problem of measuring such a device to clearly show the improvement achieved. We have made some advances in this field and these will be published in following articles.

\section{Design of feed-horns with low cross-polarization and side-lobe levels}
\label{section 7}
Satisfying the balanced hybrid condition by using a metamaterial for creating the adequate boundary condition in the inner face of a horn (\ref{equation_9}) is not the only constraint for designing a feedhorn with low levels of cross-polarization. The profile of the horn and its dimensions are determinant too. Furthermore, sometimes the sidelobe level increases whilst the cross-polarization decreases when adjusting a profile and inner dimensions or vice versa, making a feedhorn unusable since for high performance both are required to be low. The reason for this is related to the creation or disappearance of modes over the length of the horn, or the change in relative phase-shift between them because of geometrical effects. 

In order to illustrate the first constraint which is the importance of the profile, we have run calculations with the finite element method (FEM) solver of CST Microwave Studio Suite comparing the results of cross-polarization, side-lobe levels and return-loss for 4 different profiles of smooth-horn (no metamaterial or corrugations on its inner face). The horn is fed by a mono-mode guide and different ''A'' parameters have been entered in the profile curve proposed by \citet{Clarricoats} and given by

\begin{equation}
r(z)=r_{11}+(r_{12}-r_{11})\left[\frac{z}{L}(1-A)+Asin^2\left(\frac{z\pi}{2L}\right)\right] \;,
\label{equation_139}
\end{equation}

where $r_{11}$ is the input radius of the horn which adopts the value 10 mm, $r_{12}$ is the output radius which adopts the value 60 mm, $L$ is the length of the horn with value 200 mm and A is the parameter to vary. The results of the calculations are summarized in the table \ref{table_1} where it can be seen that the profile of the horn affects the behaviour and characteristics by several decibels for some cases. In our experience with corrugated horns we have found particular cases where the difference is even stronger, of the order of the tens of dBs, for cross-polarization and side-lobes levels.\newline

\begin{table}[]
\label{table_1}
\begin{tabular}{cccccccccc}

A   & \multicolumn{3}{c}{Return-Loss{[}dB{]}@{[}GHz{]}} & \multicolumn{3}{c}{Side-lobe Level{[}dB{]}@{[}GHz{]}} & \multicolumn{3}{c}{Cross-polarization{[}dB{]}@{[}GHz{]}} \\  
    & 10& 15 & 20 & 10   & 15  & 20  & 10   & 15   & 20  \\ \toprule
0   & -24         & -38         & -31.5       & -12.4         & -19.0        & -14.7        & -17.1          & -17.4         & -18.4         \\
0.3 & -26         & -36         & -31         & -12.1         & 17.2         & -14.2        & -17.5          & -17.4         & -18.4         \\
0.5 & -25         & -33.5       & -35.7       & -11.4         & -16.1        & -12.1        & -17.6          & -16.2         & -18.6         \\
1   & -27.5       & -33.5       & -31         & -9.1          & -4.2         & -8.9         & -15.5          & -15.1         & -18.9       
\end{tabular}
\caption{Influence of the profile on the behavior of a feedhorn.}
\end{table}

In order to illustrate the second constraint, the importance of the surface impedance of the inner walls of the horn, we have run calculations of the surface impedance of several metamaterials and studied their agreement with the \textit{balanced hybrid mode condition} in (\ref{equation_9}).  Once a profile is optimized, in order to create a new meta-material with a bandwidth factor over 2:1 with cross-polarization levels better than -40 dB over the entire band, i.e., satisfying the hybrid-condition, the use of novel metamaterials is necessary.  

Thus, it is necessary to be able to calculate or estimate surface impedance in order to be able to design new metamaterials with valid values for $Z_{TE,TM}$. This is not trivial. We have developed a similar theoretical framework to that presented in \cite{Lier} to correctly run calculations and simulations. This framework is summarized in the following lines.

\begin{figure}[ht!]
		\includegraphics[width=0.5\textwidth]{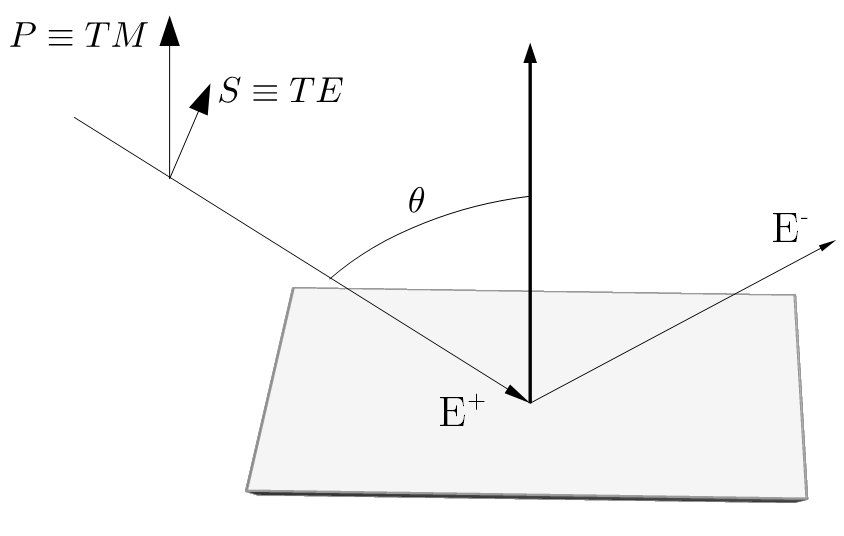}
		\centering 
		\caption{Schema of TE and TM polarization of electromagnetic modes.}
		\label{fig3_1}
	\end{figure}

It is well known that the reflection coefficient for a wave travelling through a vacuum and perpendicularly to a plane surface is given by

\begin{equation}
\Gamma=\frac{E^-}{E^+}=\frac{Z_s-Z_0}{Z_s+Z_0} \;,
\label{equation_140}
\end{equation}

where $E^+$ represents the forward incident wave and $E^-$ represents the backward reflected wave, $Z_s$ is the impedance of the plane surface and $Z_0$ is the impedance of free space. This is represented in the Fig. \ref{fig3_1}. 

Thus, it is possible to obtain expressions for TE and TM modes being reflected by a surface with characteristic $Z^{TE,TM}$ impedance. In order to obtain these expressions, it must be noted that the impedance of the free space depends on the incidence angle ($\theta$) of the forward wave. The values are, for TM-polarized modes

\begin{equation}
TM
\begin{cases} 
 |H\big|= H_0 \\
 |E\big|=E_0cos\theta
\end{cases} 
\Rightarrow Z_0^{TM}(\theta)=\frac{E^{TM}(\theta)}{H^{TM}(\theta)}=Z_0cos\theta,
\label{equation_141}
\end{equation}

while for TE-polarized modes

\begin{equation}
TE
\begin{cases} 
 |H\big|= H_0 cos\theta\\
 |E\big|=E_0
\end{cases} 
\Rightarrow Z_0^{TE}(\theta)=\frac{E^{TE}(\theta)}{H^{TE}(\theta)}=\frac{Z_0}{cos\theta}.
\label{equation_142}
\end{equation}

From equations (\ref{equation_140}), (\ref{equation_141}) and (\ref{equation_142}) it is straight forward to obtain the following expression for the surface impedance of reflected TE-modes

\begin{equation}
Z_s^{TE}=\frac{Z_0}{cos \theta} \frac{1+\Gamma^{TE}}{1-\Gamma^{TE}}\;,
\label{equation_143}
\end{equation}

and for reflected TM-modes

\begin{equation}
Z_s^{TM}=Z_0 \, cos \theta \, \frac{1+\Gamma^{TM}}{1-\Gamma^{TM}}\;.
\label{equation_144}
\end{equation}

Thus, we are able to obtain theoretical calculations of the surface impedance once we have calculated the reflection coefficient $\Gamma$ of each mode, and then introduce both reflection coefficients in (\ref{equation_9}) in order to know if our metasurface satisfies the hybrid-mode condition. This procedure can be followed by using CST Studio Suite$^{\textregistered}$. CST includes a finite element method (FEM) 3D Electromagnetic analysis software package for components and systems. In our simulations, a plane wave is launched at a grazing angle of 80 degrees at a metamaterial containing an E-field probe just above its surface. Comparing measurements of the E-field probe in the same position with and without presence of the metamaterial under test, it is possible to subtract the incoming and the reflected signals over the frequency range of interest using time gating if necessary. This allows us to obtain the reflection coefficient for each TE or TM polarization from the equations (\ref{equation_143}) and (\ref{equation_144}).

\begin{figure}[ht!]
		\includegraphics[width=0.9\textwidth]{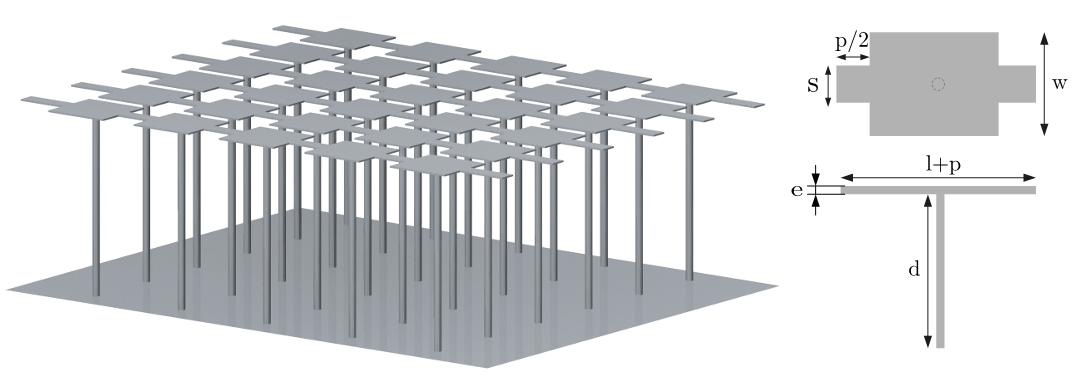}
		\centering 
		\caption{Mushroom type metamaterial based on \citet{Scarborough2}. For covering the band 10-20 GHz, the parameters can adopt the values w = 2, s = 0.4, p = 1, l = 2 and d = 5.2 mm. The thickness (e) is 0.05 mm and the diameter of the cylinders is 0.2 mm. The lattice period or space between consecutive lines is 0.4 mm.}
		\label{fig3_2}
	\end{figure}

\begin{figure}[ht!]
		\includegraphics[width=0.4\textwidth]{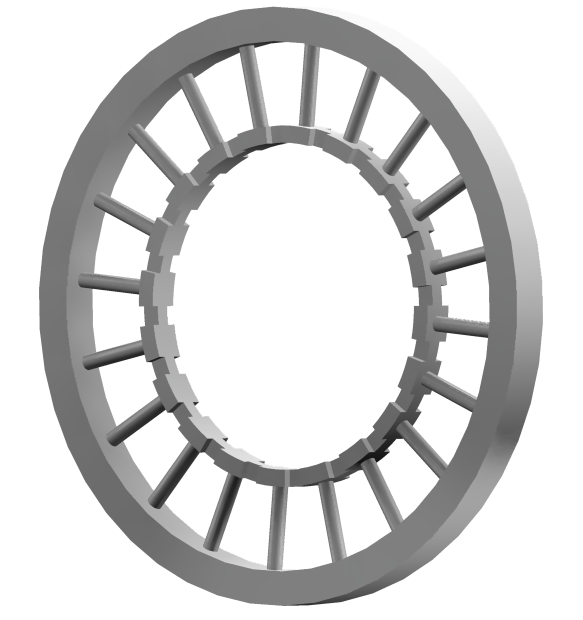}
		\centering 
		\caption{Ring of the meta-horn antenna. Here the more realistic dimensions of the model-2 are taken in a ring of 17.4 mm exterior radius.}
		\label{fig3_3}
	\end{figure}

These simulations have been taken for a mushroom 3D-type metamaterial based on \citet{Scarborough2} presented in fig.\ref{fig3_2}.  This metamaterial is named as \textit{model-1}. It is necessary to bear in mind that the aim of these simulations is to design a practical inner surface of a horn-antenna. If we apply this here we see that \textit{model-1} could be manufactured as a set of rings each with one line of the metasurface pattern such as the shown in Fig. \ref{fig3_3}. 

\begin{figure}[ht!]
		\includegraphics[width=0.8\textwidth]{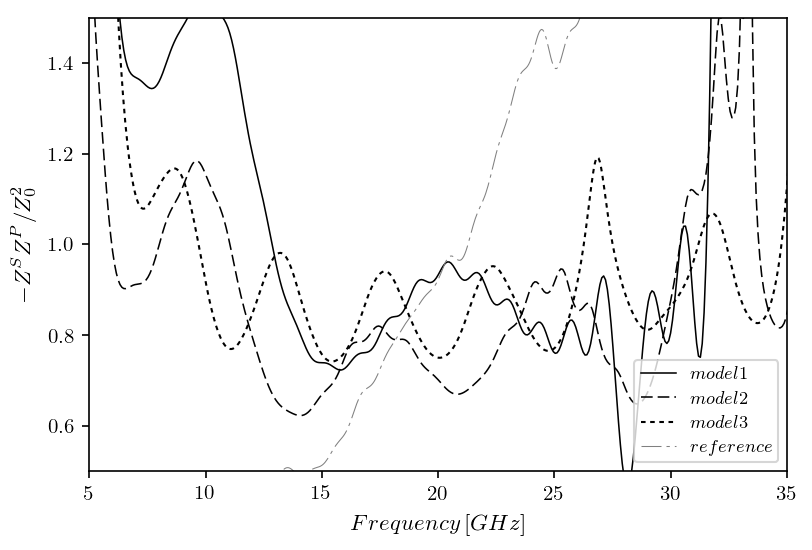}
		\centering 
		\caption{Metamaterial FEM-simulations for 3 different models. The metamaterial satisfies the hybrid mode condition when {$-Z^{S} Z^{P}/Z_0^2$} is exactly 1. The solid line represents the results for the model-1 mushroom topology metamaterial inspired by \citet{Scarborough2}. The dashed line corresponds to the modified model-1 to a more realistic design, model-2 mushroom metamaterial, with diameter of cylinder = 0.85, s = 0.9 and thickness of the plates equal 1 mm. The dotted line represents both top-plated and inverted-"L" topologies of model-3, since they presents identical results. The dot-stripes gray line represents an equivalent model using corrugations instead of mushrooms and is used as a reference. The modelled material is perfect electric conductor (PEC) for all simulations.}
		\label{fig3_4}
	\end{figure}

From the hybrid-mode condition in \ref{equation_9},  a metamaterial satisfies the hybrid condition when $-Z^{TE} Z^{TM}/Z_0^2$ is exactly 1 (see Fig. \ref{fig3_4}). Thus, a metamaterial with $-Z^{TE} Z^{TM}/Z_0^2=1$ would theoretically lead inner walls for feedhorn antennas with null cross-polarization levels, while it is expected that values of $-Z^{TE} Z^{TM}/Z_0^2$ in the vicinity of 1 yield $>-20 dB$ of cross-polarization attenuation. This is the case for this metamaterial, whose dimensions are given in Fig. \ref{fig3_2} and whose simulation will be represented later by the solid black line of the Fig. \ref{fig3_4}, presenting a correct behaviour in the band 12-28 GHz approximately which is approximately a 2.5:1 bandwidth factor.

However, the small dimensions for the sample of metamaterial in figure \ref{fig3_2} make it impossible to fabricate with either workshop machining or state-of-the-art additive techniques. Thus, we have established several constraints. Firstly, the 0.05 mm thickness is far from practically rigid structure. After several tests, a conservative value for this thickness has been established as 1 mm\footnote{This values is 20 times thicker than the original design, which is too thin to be manufactured. This varies the surface capacitance and inductance as well as the effective depth of the groove with an effect shown in Fig. \ref{fig3_4} and discussed in the text.}. On the other hand, the diameter of the cylinder (i.e, the pin or base of the plates) is difficult to manufacture by these techniques, so it has been set at 0.85 mm. Thirdly, the dimension of the parameter "s" has been increased to 0.9 mm. The rest of parameters for model-2 have the same value as model-1.\newline Now is possible to compare the behaviour in terms of hybrid mode condition of model-1 and its modification yielding model-2. The effect of these modifications to the model-1 design with the original dimensions based on \citet{Scarborough2} can be compared with the help of the results of the new simulation shown in Fig. \ref{fig3_4}, where  this topology is named as \textit{model-2}, as said. It is possible to see that there is not a significant change in the bandwidth, the value $-Z^{TE} Z^{TM}/Z_0^2\sim1$ being slightly worse for midband frequencies but slightly increasing the bandwidth to lower frequencies. The explanation of this increment of bandwidth could be due to the benefit in terms of capacitive impedance given by the increase in cross section of the plates and cylinders while the inductance is slightly reduced.

\begin{figure}[ht!]
		\includegraphics[width=0.6\textwidth]{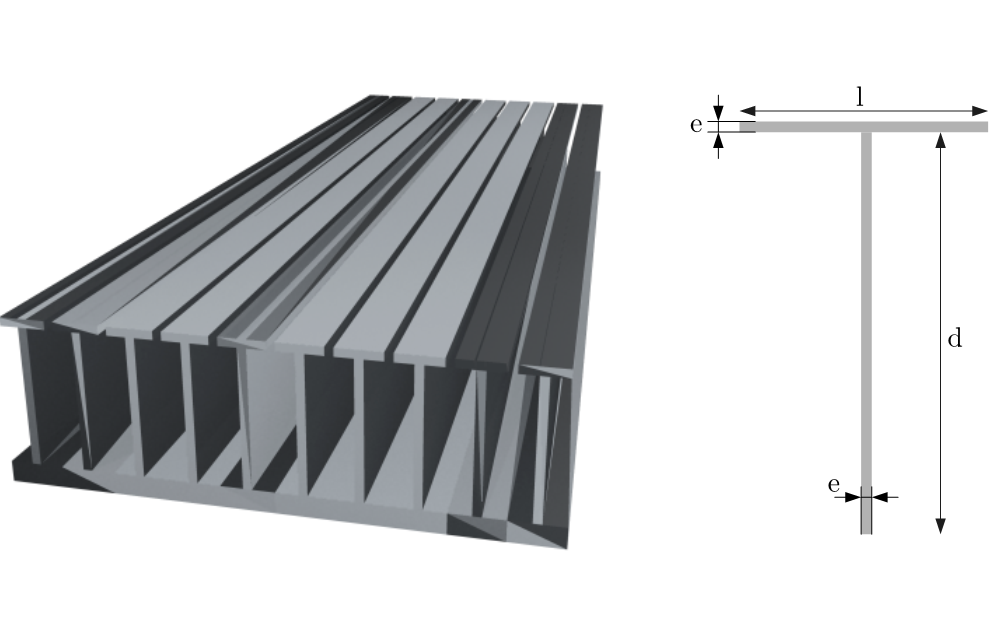}
		\centering 
		\caption{Top-plated corrugations in a simplified version of the original metamaterial where e=0.4, l=1.66 d=6.5 mm and the free space between lines is 0.25 mm. Note that a standard mechanization allows a reduction in the thickness (e).}
		\label{fig3_5}
	\end{figure}

We want to compare the highly-complex topology of metamaterials based on \citet{Scarborough2} (model-1) and its more realistic version (model-2) with a simpler version redesigned in order to be manufactured by traditional mechanical workshop techniques. In order to do this, the cylindrical pattern is reduced to a solid wall, and the top plates are also unified, forming a "tee" from a frontal view. This design is presented in Fig. \ref{fig3_5} with the name of \textit{top-plated corrugations}. 

\begin{figure}[ht!]
		\includegraphics[width=0.6\textwidth]{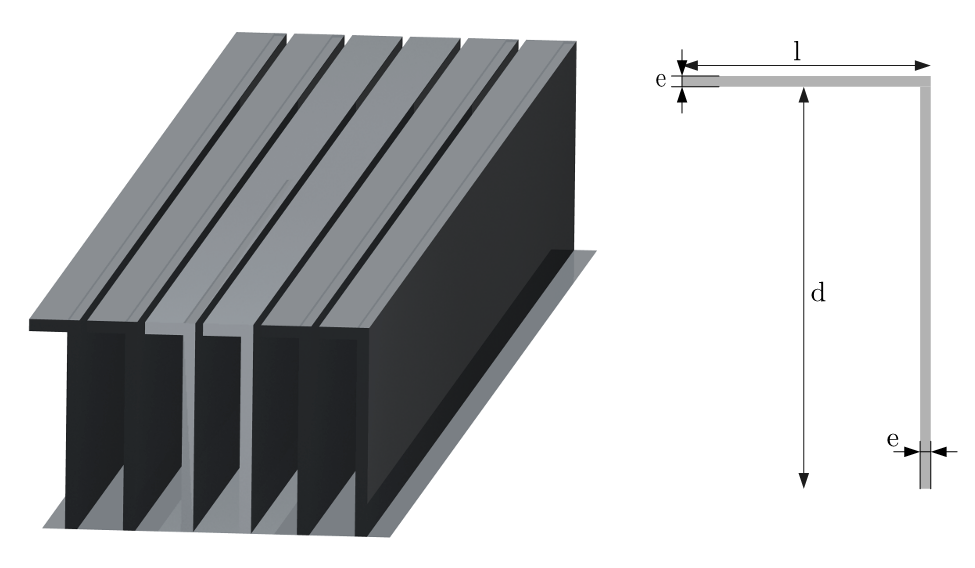}
		\centering 
		\caption{Inverted "L" in a simplified version of the original metamaterial where e = 0.4, l = 1.66, d = 6.5 mm and the lattice free space between lines is 0.25 mm.}
		\label{fig3_6}
	\end{figure}

\begin{figure}[ht!]
		\includegraphics[width=0.6\textwidth]{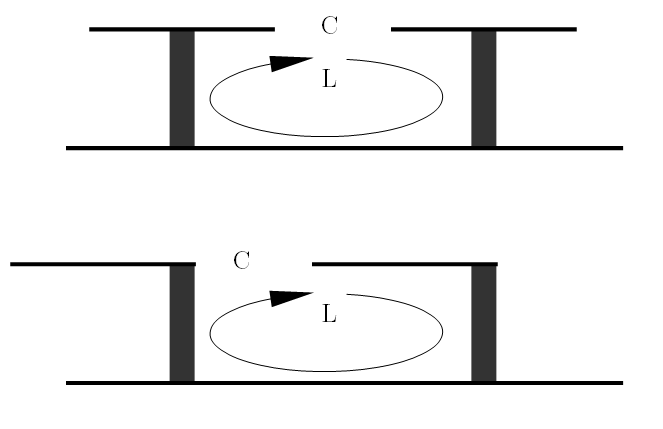}
		\centering 
		\caption{To show top-plated (upper) and inverted-"L" (lower) topologies similarity in terms of impedance. Since $R \simeq 0$ for good conductors, a generic impedance approximates the value $Z=j\omega L-j/(\omega C)$ in both models, where $L$ and $C$ are given values of inductance and capacitance respectively. }
		\label{fig3_7}
	\end{figure}

Moreover, we reconvert the top-plated topology presented in Fig. \ref{fig3_5} to adapt it to a new thin version of the classic ring-loaded corrugation topology \cite{Clarricoats}, more frequently used in taper-converters for feedhorns. This thin-ring-loaded or inverted-"L" topology is presented in  Fig. \ref{fig3_6}. Simulations of this inverted-"L" metamaterial indicate that both top-plated and inverted-"L" topologies are equivalent from an electromagnetic point of view, so both could be combined or used for different applications depending on the constraints and requirements of each case. This electromagnetic similarity is logical because both topologies have identical surface inductance and capacitance. This idea can be qualitatively understood with the help of the Fig. \ref{fig3_7}. The dotted black line in the Fig. \ref{fig3_4} named as model-3 represents both top-plated and inverted-"L" topologies. This figure shows that the bandwidth has improved for this simplified metamaterial model-3. We think that this is for several reasons; firstly, the reduction in the limiting thickness of the plates (adopting now the value 0.4 mm for every thickness)\footnote{Note that mechanical workshop techniques allow thinner dimensions than in model-2, designed for the more restrictive additive printing.}, but mainly because of the gain in spatial symmetry in comparison with the metamaterial of models 1 and 2, where different points in the space between lines have different values of surface impedance, depending on the proximity to different geometrical components. This has been confirmed by spatially displacing the electric field probe in different simulations. Moreover, arbitrary points between plates arbitrary points between plates of width $w$ and plates of width $s$ in the top of the mushroom structure of the model Fig. \ref{fig3_2} have different surface impedance, as expected due to the difference in capacitance because of the differential space between plates. 

Thus, from the analysis of Fig. \ref{fig3_4}, it can be seen that the operating bandwidth is theoretically even wider for the top-plated and inverted-"L" corrugation topology (model-3), covering the 6.5-35 GHz band for an impressive 5:1 bandwidth factor. Also the value of $-Z^{S} Z^{P}/Z_0^2$ is closer to the unity over the whole band, theoretically presenting a better behavior in terms of cross polarization attenuation and satisfaction of the hybrid mode condition. It can be concluded that, at least in theory, the metamaterial presented in \citet{Scarborough2} does not have any real advantage over a simpler and "easily" manufacturable thin version of the ring-loaded topology, presented around 30 years ago by several authors (e.g., \cite{Clarricoats}) and now presented here as a metamaterial. 

Finally, it is interesting to conclude this analysis comparing these new metamaterials with the most extensively used metamaterial  for feed-horns these days, which are corrugations, taking the corrugations as a reference. This reference helps to see how innovative these new metamaterials are. Thus, an equivalent version of a corrugated surface has been simulated in order to see its bandwidth. In this reference model, the depth of the slots is 6.9 mm, the thickness of the corrugations is 0.4 mm and the lattice period or space between two consecutive lines of corrugations is an optimized value around 1.5 mm. The result of the simulations of this model is represented by the dot-stripes gray line in Fig. \ref{fig3_4}, where it can be seen that the product of normalized impedance has a big slope so that its bandwidth factor is much lower than for the metamaterial-based models, being around 1.4:1  or a 40\% bandwidth, as the bibliography predicts.

\section{Conclusions}
\begin{figure}[ht!]
		\includegraphics[width=1\textwidth]{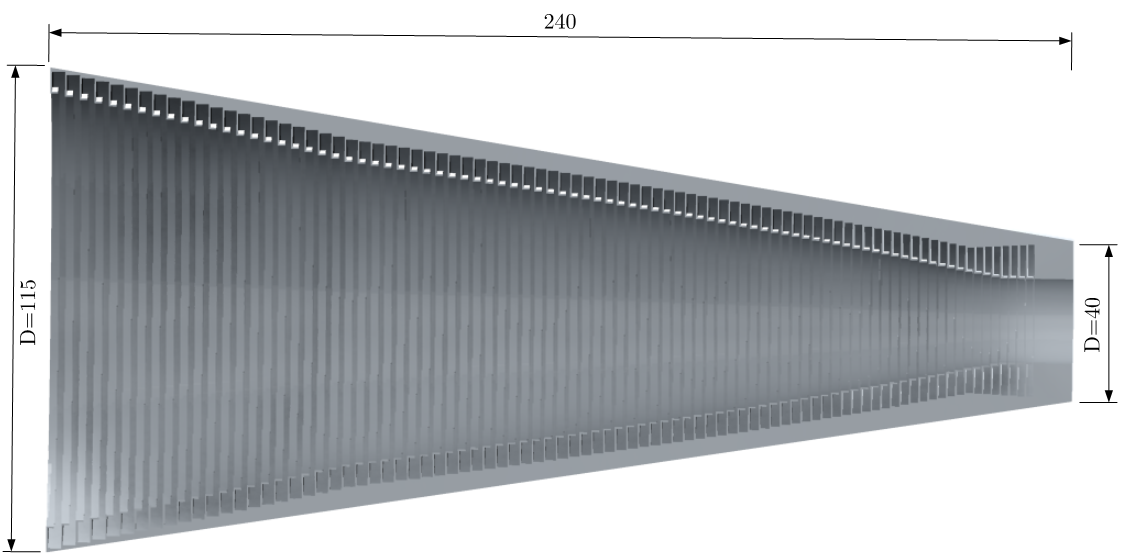}
		\centering 
		\caption{Design of meta-horn. General dimensions, in millimeters.}
		\label{fig17}
	\end{figure}

\begin{figure}[ht!]
		\includegraphics[width=0.8\textwidth]{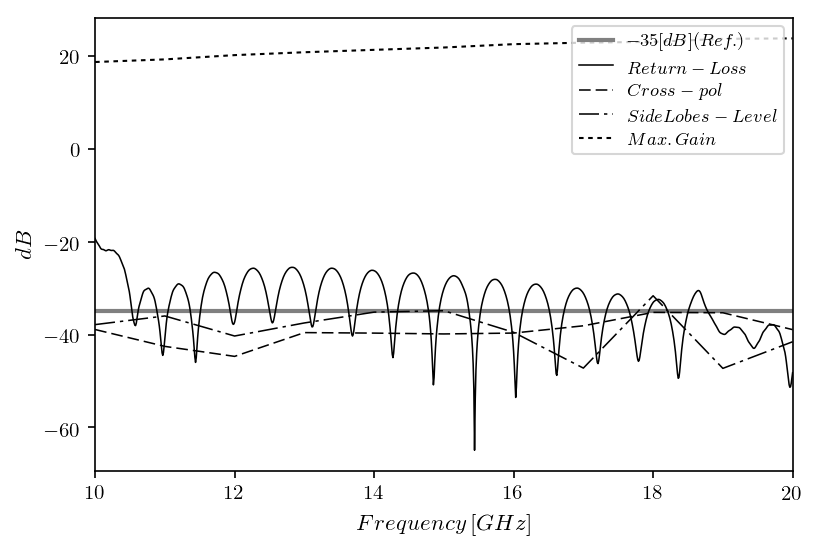}
		\centering 
		\caption{Design of meta-horn. Results of the simulations.}
		\label{fig18}
	\end{figure}

In this work, we presented the fundamentals to achieve low cross polarization levels in feedhorns. Now we will show a practical-theoretical example using the metamaterial of Fig. \ref{fig3_6} to design a conical horn composed of meta-rings, which can be manufactured and coupled, for example, with the technique published in \cite{Milano}. The design of the horn and the very promising results based on CST Studio Suite$^{\textregistered}$ simulations are shown in Figs. \ref{fig17} and \ref{fig18}. In Fig. \ref{fig18} an excellent behaviour over an entire 2:1 bandwidth factor band in terms of return-loss, side-lobe levels, gain and cross-polarization, with levels better than -35 dB over the entire 10-20 GHz band, is predicted. 

This design is very compact even using a conical profile because easy and quick fabrication was desired, but the benefits could be improved, as we have already explained in section \ref{Section_6.2}, by choosing an adequate profile.

\section*{Acknowledgment}

The authors wish to acknowledge the support of all the technicians, engineers, scientists and administrative staff of the IAC and Quijote-CMB Experiment.

\newpage

\begin{appendices}
\setcounter{equation}{0}
\setcounter{figure}{0}
\renewcommand\theequation{A.\arabic{equation}}
\renewcommand\thefigure{\thesection.\arabic{figure}}  

\section{\\Electromagnetic Field through Smooth Cylindrical Waveguides}

\label{Appendix-A}

The first topic in microwave physics or engineering is the resolution of the Maxwell's equations for the particular case

\begin{equation}
{\vec{\nabla}}\cdot {\vec  {D}}=\rho  \;
\label{equation_A1}
\end{equation}

\begin{equation}
\vec{\nabla} \cdot \vec{B} = 0  \;
\label{equation_A2}
\end{equation}

\begin{equation}
\vec{\nabla} \times \vec{E} = - \frac{\partial{\vec{B}}}{\partial{t}}  \;
\label{equation_A3}
\end{equation}

\begin{equation}
\vec{\nabla} \times \vec{H} = \vec{J} + \frac{\partial{\vec{D}}}{\partial{t}}  \;.
\label{equation_A4}
\end{equation}

Equation (\ref{equation_A1}) is Gauss's Law, (\ref{equation_A2}) is known as Gauss's Law for Magnetism, (\ref{equation_A3}) is the Maxwell-Faraday Equation and equation (\ref{equation_A4}) is the Ampere's Circuital Law. The nabla symbol, $\nabla$, denotes the three-dimensional gradient operator. $ E $ denotes the electric field (usually in V/m in the international system of units), $ H $ the magnetic field intensity [A/m], $ D $ the electric flux density [C/$ m ^ 2 $], $B$ the magnetic field flux [Wb/m or Tesla, T], $J$ the current density [A/$m^2$] and $\rho $ the electric charge density [C/$ m ^ 3 $]. The universal constants implicit in the equations are $c=1/\sqrt{\varepsilon_{0} \mu_0}$ the speed of light in the vacuum [m/s], $\varepsilon_0$ [F/m] the permittivity of the vacuum and $\mu_0$ [H/m] is the permeability of free space (i.e., vacuum). 

The derivation of the specific decoupled solution from the Maxwell equations for waveguides is well known. It can be shown that

\begin{equation}
\nabla_t^{2} E_z + (k^2+\gamma^2) E_z=0 \;
\label{equation_A5}
\end{equation}

\begin{equation}
\nabla_t^{2} \vec{E_t} + (k^2+\gamma^2) \vec{E_t}=0 \;
\label{equation_A6}
\end{equation}

\begin{equation}
\nabla_t^{2} H_z + (k^2+\gamma^2) Hz=0 \;
\label{equation_A7}
\end{equation}

\begin{equation}
\nabla_t^{2} \vec{H_t} + (k^2+\gamma^2) \vec{H_t}=0   \;,
\label{equation_A8}
\end{equation}

where the suffix $t$ denotes transverse waves and the suffix $z$ denotes longitudinal waves, $k=\omega \sqrt{\mu \varepsilon}$ and the complex propagation constant\footnote{${\frac  {A_{0}}{A_{z}}}=e^{{\gamma z}}$ and $\gamma=\alpha +i\,\beta$, where $\alpha$ is the attenuation constant and $\beta$ is the phase constant.}$\gamma$ is unknown. Only the scalar equations (\ref{equation_A5}) and (\ref{equation_A7}) have to be solved because the equations (\ref{equation_A9}) and (\ref{equation_A10}) shown below are satisfied.

\begin{equation}
\vec{E_t}=\frac{1}{k^2+\gamma^2} \left(-j\omega \mu \nabla_t \times H_z \vec{u}_z - \gamma \nabla_t E_z \right)  \;
\label{equation_A9}
\end{equation}

\begin{equation}
\vec{H_t}=\frac{1}{k^2+\gamma^2} \left(j\omega \nabla_t \times E_z \vec{u}_z - \gamma \nabla_t H_z \right)  \;.
\label{equation_A10}
\end{equation}

Equations (\ref{equation_A9}) and (\ref{equation_A10}) refer to the case of progressive waves. For the case of regressive waves, the sign of $\gamma$ is inverted.

\begin{figure}[ht!]
		\includegraphics[width=0.6\textwidth]{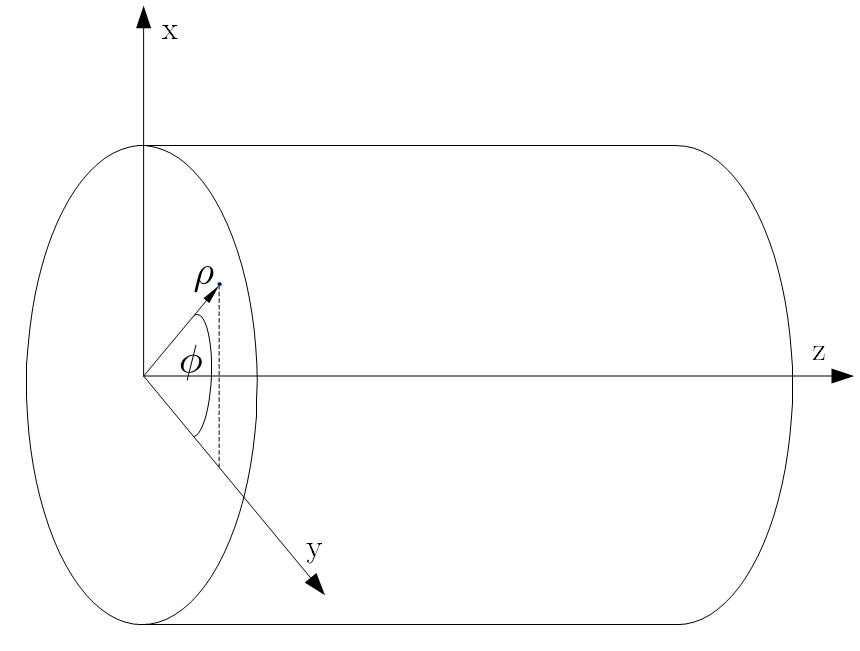}
		\centering 
		\caption{Cylindrical waveguide.}
		\label{figA_1}
	\end{figure}
 
For the particular case of cylindrical waveguides we start solving the wave equations (\ref{equation_A5}) and (\ref{equation_A7}) after the transformation $k_c=k^2+\gamma^2$ and assuming $A_z$ equal to $E_z$ or $H_z$ indistinctly for simplicity since the derivations are equivalent in both cases. This is
 
\begin{equation}
\nabla_t^{2} A_z + k_c \, A_z=0 \;.
\label{equation_A11}
\end{equation}

This leads to the following differential equation with partial derivatives (previously transformed to cylindrical coordinates $z$, $\rho$ and $\phi$)

\begin{equation}
\frac{\partial^{2}A_z}{\partial \rho^{2}}+\frac{1}{\rho}\frac{\partial A_z}{\partial \rho}+\frac{1}{\rho^{2}}\frac{\partial^{2}A_z}{\partial \phi^{2}}+k_c^{2} \, A_z\;.
\label{equation_A12}
\end{equation}

The equation (\ref{equation_A12}) can be solved by using the separation of variables method. This implies the transformation

\begin{equation}
A_z(\rho,\phi)=R(\rho) \cdot F(\phi) \;.
\label{equation_A13}
\end{equation}

This separation of variables leads to the following differential equation, where the apostrophe denotes the order of the ordinary derivative

\begin{equation}
\rho^2 \, \frac{R''}{R} + \rho \, \frac{R'}{R}+\frac{F''}{F} + \rho^2 \, k_c^2=0 \;,
\label{equation_A14}
\end{equation}

and, since the sum of the derivatives of both variables are constant values (equal to 0 in this case), separating the equation above we obtain two independent ordinary differential equations with well known solutions, where the term $\xi$ represents an arbitrary constant

\begin{equation}
\rho^2 \, \frac{R''}{R} + \rho \, \frac{R'}{R} + \rho^2 \, k_c^2= \xi^2 \;
\label{equation_A15}
\end{equation}

\begin{equation}
\frac{F''}{F}=-\xi^2 \;.
\label{equation_A16}
\end{equation}

The equation (\ref{equation_A15}) is the Bessel Differential Equation, while the equation (\ref{equation_A16}) is a regular exponential differential equation, with a known solution. Thus, the solution of (\ref{equation_A15}) is given by 

\begin{equation}
R=C \cdot J_{\xi}(k_c\rho)+D \cdot N_{\xi}(k_c\rho) \;.
\label{equation_A17}
\end{equation}
In this equation, C and D are constants, $J_{\xi}$ represents a Bessel function of the first kind of order $\xi$ and $N_{\xi}$ is a Bessel function of the second kind or Neumann function of order $\xi$ . Since all the Bessel functions have a singularity at the origin ($\rho = 0$) except those of the first kind, the final solution of (\ref{equation_A15}) must be given by ($D=0$)

\begin{equation}
R=C \cdot J_{\xi}(k_c\rho) \;,
\label{equation_A18}
\end{equation}

while the solution of (\ref{equation_A16}) is

\begin{equation}
F=A \, cos \, (\xi \phi) + B \, sin \, (\xi \phi)   \;,
\label{equation_A19}
\end{equation}

where A and B are constants.

This yields the solution of the equation (\ref{equation_A12}), given by

\begin{equation}
A_z=J_{\xi}(k_c \rho) [A \, cos \, (\xi \phi) + B \, sin \, (\xi \phi)]   \;.
\label{equation_A20}
\end{equation}

Since $\phi$ is periodic ($\phi \in [0,2\pi]$), $\xi$ must be an integer number $n$. This leads to the final solution

\begin{equation}
A_z=J_{n}(k_c \rho) [A \, cos \, (n \phi) + B \, sin \, (n \phi)]   \;.
\label{equation_A21}
\end{equation}

It is interesting now to study the individual cases for TM and TE waves starting from the general solution (\ref{equation_A21}).

\subsubsection{Transverse Magnetic mode case}

For the particular case of a transverse magnetic (TM) wave we have $H_z=0$  and  $E_z \neq 0$. 
\newline At the surface of the cylindrical guide $\rho = a$ and $E(z)_{\rho=a}=0$, and thus we obtain

\begin{equation}
J_{n}(k_c \, a) = 0   \;,
\label{equation_A22}
\end{equation}

and 

\begin{equation}
k_c \, a = P_{n \, l} \rightarrow k_c= \frac{P_{n \, l}}{a}  \;,
\label{equation_A23}
\end{equation}

where $P_{n \, l}$ is the $l$-th root of $J_n$. We will have occupation of $TM_{n \, l}$ modes, where $n$ comes from the factor $\xi$ used in the inverse derivative and $l$ is derived from $k_c= \frac{P_{n \, l}}{a}$.

Now the longitudinal field equation for this particular case can be written using eq. (\ref{equation_A21})

\begin{equation}
E_z=J_{n}(k_c \rho) \, [A \, cos \, (n \phi) + B \, sin \, (n \phi)]  \;,
\label{equation_A24}
\end{equation}

and the tangential (cylindrical) components from equations (\ref{equation_A9}) and (\ref{equation_A10}) are

\begin{equation}
E_{\rho}=-j\frac{\beta}{k_c} \, J_n'(k_c \rho) \, [A \, cos \, (n \phi) + B \, sin \, (n \phi)] 
\label{equation_A25}
\end{equation}

\begin{equation}
E_{\phi}=-j \frac{\beta \, n}{k_c^2 \, \rho} \, J_n (k_c \rho) \, [B \, cos \, (n \phi) - A \, sin \, (n \phi)]  \; ,
\label{equation_A26}
\end{equation}

where the phase-term $e^{-j \beta z}$ is omitted in both equations and the propagation constant of a mode $TM_{n \, l}$ is

\begin{equation}
\beta_{n\,l} = \sqrt{k^2 - k_c^2}=\sqrt{k^2-(P_{n\,l}/a)^2} \; .
\label{equation_A27}
\end{equation}

For the calculation of the magnetic field, the concept of wave impedance can be applied\footnote{Equivalent concept to the surface impedance $Z_s$ applied later in this chapter, so the derivation can be seen there.}, and it can be easily calculated from the equations

\begin{equation}
H_{\rho}=-\frac{E_{\phi}}{Z^{TM}}
\label{equation_A28}
\end{equation}

\begin{equation}
H_{\phi}=\frac{E_{\rho}}{Z^{TM}} \; ,
\label{equation_A29}
\end{equation}

where the wave impedance is $Z^{TM}=\frac{\beta}{\omega \varepsilon}$ and the cut-off frequency for the $TM_{n \, l}$ mode is

\begin{equation}
f_c=\frac{k_c}{2 \, \pi \sqrt{\mu \varepsilon}}=\frac{P_{n \, l}}{2 \, \pi \,a \sqrt{\mu \varepsilon}} \; ,
\label{equation_A30}
\end{equation}

and the mode with the lower cutoff frequency will be given by the first zero of the Bessel function $P_{0 \, 1} \rightarrow TM_{0 \, 1}$. In spite of this, the cut-off frequency on a circular waveguide will be given by a $TE$ mode, as we will see in \ref{TEM}.

\subsubsection{Transverse Electric mode case} \label{TEM}

For the particular case of a transverse electric (TE) wave we have $E_z=0$  and  $H_z \neq 0$. 
\newline The tangential field to the waveguide will be $E_{tan}=0$ only if

\begin{equation}
\frac{\partial{H_z}}{\partial{\rho}} \bigg\vert_{\rho=a}=0 \; .
\label{equation_A31}
\end{equation}

The magnetic longitudinal field equation for this particular case can be written using eq. (\ref{equation_A21})

\begin{equation}
H_z=J_{n}(k_c \rho) \, [A \, cos \, (n \phi) + B \, sin \, (n \phi)]  \;,
\label{equation_A32}
\end{equation}
and thus,

\begin{equation}
J'_n(k_c \, a)=0 \; ,
\label{equation_A33}
\end{equation}

where $J'_n$ is derived from the Bessel function. In an analogous way to the analysis completed for the TM case, here it is possible to obtain

\begin{equation}
k_c=\frac{P'_{n\, l}}{a} \; ,
\label{equation_A34}
\end{equation}

where $n$ refers to the number of circumferential ($\phi$) modes and $l$ refers to the number of radial ($\rho$) modes. The tangential (cylindrical) components, from equations (\ref{equation_A9}) and (\ref{equation_A10}), are given by

\begin{equation}
H_{\rho}=-j\frac{\beta}{k_c} \, J_n'(k_c \rho) \, [A \, cos \, (n \phi) + B \, sin \, (n \phi)] 
\label{equation_A35}
\end{equation}

\begin{equation}
H_{\phi}=-j \frac{\beta \, n}{k_c^2 \, \rho} \, J_n (k_c \rho) \, [B \, cos \, (n \phi) - A \, sin \, (n \phi)]  \; ,
\label{equation_A36}
\end{equation}

where the propagation constant of a $TE_{n \, l}$ mode is
\begin{equation}
 \beta_{n\,l} = \sqrt{k^2 - k_c^2}=\sqrt{k^2-(P'_{n\,l}/a)^2} \; ,
\label{equation_A37}
\end{equation}
and where the phase-term $e^{-j \beta z}$ has been omitted for clarity.

Analogously to the TM case, the concept of wave impedance can be applied in order to obtain the expressions for the electric field

\begin{equation}
H_{\rho}=E_{\phi} \, Z^{TE}
\label{equation_A38}
\end{equation}

\begin{equation}
E_{\phi}=-H_{\rho} \, Z^{TE} \; ,
\label{equation_A39}
\end{equation}

where the wave impedance is $Z^{TE}=\frac{\omega \mu}{\beta}$ and the cut-off frequency for the $TE_{n \, l}$ mode is

\begin{equation}
f_c=\frac{k_c}{2 \, \pi \sqrt{\mu \varepsilon}}=\frac{P'_{n \, l}}{2 \, \pi \,a \sqrt{\mu \varepsilon}} \; ,
\label{equation_A40}
\end{equation}

so the mode with the lower cutoff frequency will be given by the first zero of the derived Bessel function $P'_{1 \, 1} \rightarrow TE_{1 \, 1}$. 

Since $P'_{1 \, 1} \simeq 1.8 < P_{0,1} \simeq 2.4$ and comparing the equations (\ref{equation_A30}) and (\ref{equation_A40}), it is possible to conclude that the $TE_{1\, 1}$ is the fundamental mode (has a lower cuttoff frequency) in a circular waveguide. Then, the $TE_{1\, 1}$ can be theoretically excited in the absence of higher order modes (in fact, because of the well known values of the zeros of the Bessel functions and their derivatives, it is possible to conclude now that the  $TE_{1\, 1}$ and the $TM_{0 \, 1}$ are the first two propagating modes) in a circular waveguide.

Now, it is worth noting that a smooth metallic cylindrical waveguide could never satisfy the hybrid mode condition expressed in equation (\ref{equation_9}). From the analysis shown in this appendix it is straightforward to derive the following expression

\begin{equation}
Z^{TE}Z^{TM}=\frac{\beta}{\omega \varepsilon} \frac{\omega \mu}{\beta}=\frac{\mu}{\varepsilon}\neq \frac{\mu_0}{\varepsilon_0}.
\label{equation_A41}
\end{equation}

Thus, in \ref{equation_A41}, and since $\mu_0/\varepsilon_0=Z_0^2$, it can be concluded that the hybryd mode condition can not be satisfied by a smooth cylindrical waveguide because its surface impedance is given by the characteristics of the metal. Thus, a metamaterial is needed in order to modify the surface impedance of the inner face of the guide.
In order to obtain a cylindrical waveguide which can satisfy the hybrid mode equation in (\ref{equation_9}), \cite{1966ITAP...14..654M} proposed the use of corrugations in the inner face of the cylindrical guide. The field equations and their relation with the hybrid mode condition are described in the Appendix-B\label{Appendix-B}.

\section{\\Electromagnetic Field through Corrugated Cylindrical Waveguides}
\setcounter{equation}{0}
\setcounter{figure}{0}
\renewcommand\theequation{B.\arabic{equation}}
\renewcommand\thefigure{\thesection.\arabic{figure}}  

\label{Appendix-B}

For the case of a corrugated waveguide, the simple solution found for the propagation function in the appendix-\ref{Appendix-A} does not exist. There is, however, an important exception, the case of big aperture radius ($a$) compared to the wavelength ($\lambda$). With this condition, two expressions are enough to describe the propagation function of almost all the modes

\begin{equation}
 \beta_{0\,l}=\sqrt{(ka)^2-P_{0\,l}^2}\;
\label{equation_B1}
\end{equation}

\begin{equation}
 \beta_{2\,l}=\sqrt{(ka)^2-P_{2\,l}^2}\;,
\label{equation_B2}
\end{equation}

where $P_{n \, l}$ is the $l$-th root of $J_n$ ($J_0(P_{0\,l})=0$ and $J_2(P_{2\,l})=0$).

An anisotropic surface can be obtained by radially corrugating a cylindrical waveguide\footnote{For a finite number of grooves per wavelength, the effective depth of the corrugation is modified, and a correction will be needed [\citet{Dragone2} pp. 869-888]}, obtaining the guide represented in the Fig. \ref{fig2}. Assuming $cos \phi$ variations, for a large cylinder in the $\vec{u}_z$ direction, and for an aperture size large enough\footnote{If $k\,a \rightarrow \infty$ is not satisfied, the properties of a mode depend on the surface reactance ($X_s$) of the corrugations and cross polarization exists, unless if $X_s=\infty$ and $\beta$ is not given by (\ref{equation_B1}) and (\ref{equation_B2}) \cite{Mac}.} in terms of  wavelength, the field inside a corrugated waveguide ($r\le a$) can be expressed as a superposition of the circular waveguide TE and TM modes. Analogously to (\ref{equation_B2}), it is possible to write a condition which the longitudinal electric and magnetic fields (both denoted indistinctly by $A$) satisfy

\begin{equation}
\nabla_t^{2} A_z + K_N^2 \, A_z=0 \;,
\label{equation_B3}
\end{equation}

where 

\begin{equation}
K_N^2 +\beta_N^2=k^2 \;,
\label{equation_B4}
\end{equation}

 $k=\sqrt{\varepsilon_0\mu_0}=\frac{2\pi}{\lambda}$ being the wave number and the propagation coefficient $\beta$ can be expanded in space harmonics related to the spatial lattice period of the grooves and denoted by $p$ in Fig. \ref{fig3}
\begin{equation}
\beta_N=\beta_0+\frac{2\pi N}{p} \;.
\label{equation_B5}
\end{equation}

In order to calculate the propagation coefficient $\beta_N$, the boundary condition $r=a$ is applied. When the aperture of the horn ($a$) is much larger than the wavelength ($\lambda$), the terms of $\beta_N$ for $N\ge1$ in (\ref{equation_B5}) can be ignored, transforming the propagation function into a constant value $\beta$. It can be assumed  that the number of corrugations increase if the thickness of each corrugation decreases, if the condition $b<\lambda/2$ is imposed\footnote{Note that this affects the $H_z/E_z$ ratio, altering the optimal working frequency.}.

Thus, for an equivalent transmission line circuit, the reactance at the surface defined by $r=a$ is given by

\begin{equation}
jX=jZ_0\,tan(k\,d)\;
\label{equation_B6}
\end{equation}

\begin{equation}
jX_s=jX \left(1-\frac{t}{p} \right)\;,
\label{equation_B7}
\end{equation}

where $Z_0=\sqrt{(\mu_0/\varepsilon_0)}$ is the vacuum impedance, $d$ is the depth of a groove and $t$ is its thickness.

Thus, the following boundary conditions are true for the metallic surface defined by the inner radius

\begin{equation}
r=a
\begin{cases} 
 E_{\phi} \simeq 0\\
 H_{\phi}\simeq -\frac{E_z}{j\,X_s}\;.
\end{cases} 
\label{equation_B8}
\end{equation}

The propagation constant $\beta$ in the $\vec{u}_z$ direction is given by

\begin{equation}
\beta=k\, cos\theta_1\;,
\label{equation_B9}
\end{equation}

where $\theta_1$ is real so $\beta<k$. Assuming that the $\phi$ dependence of $E_z$ is given by $cos\phi$, the field expressions from (\ref{equation_B3}) for a mode propagating along the $z$ axis with a $\beta$ propagation constant are

\begin{equation}
E_z=A\,J_1(\kappa\,r)\,cos\phi\,e^{-j\beta z} 
\label{equation_B10}
\end{equation}

\begin{equation}
H_z=\frac{1}{Z_0} B\,J_1(\kappa\,r)\,sin\phi\,e^{-j\beta z} 
\label{equation_B11}
\end{equation}

\begin{equation}
E_{\phi}=\frac{j}{sin\theta_1} \left[B\,J_1' (\kappa\,r)+A \, cos \theta_1 \frac{J_1(\kappa\,r)}{\kappa\,r}\right]sin\phi\,e^{-j\beta z}
\label{equation_B12}
\end{equation}

\begin{equation}
H_{\phi}=\frac{1}{Z_0} \frac{-j}{sin\theta_1} \left[A\,J_1' (\kappa\,r)+B \, cos \theta_1 \frac{J_1(\kappa\,r)}{\kappa\,r}\right]cos\phi\,e^{-j\beta z}
\label{equation_B13}
\end{equation}

\begin{equation}
E_{r}=\frac{-j}{sin\theta_1} \left[B\frac{J_1 (\kappa\,r)}{\kappa\,r}+A \, cos \theta_1 \, J_1' (\kappa\,r)\right]cos\phi\,e^{-j\beta z}
\label{equation_B14}
\end{equation}

\begin{equation}
H_{r}=\frac{1}{Z_0}\frac{-j}{sin\theta_1} \left[A\frac{J_1 (\kappa\,r)}{\kappa\,r}+B \, cos \theta_1 \, J_1' (\kappa\,r)\right]sin\phi\,e^{-j\beta z} \, ,
\label{equation_B15}
\end{equation}

with $r\leq a$ and where $\kappa=k\, sin\theta_1$ and $\kappa^2+\beta^2=k^2$.\newline

The introduction of the equations (\ref{equation_B10}) to (\ref{equation_B15}) into the conditions in (\ref{equation_B8}) yields

\begin{equation}
\gamma \equiv \frac{A}{B}=-\frac{u}{cos\theta_1}\frac{J_1' (u)}{J_1(u)} \, 
\label{equation_B16}
\end{equation}

\begin{equation}
y\equiv-\frac{Z_0}{X_s}=\frac{cos\theta_1}{sin\theta_1 \, u}\frac{1}{\gamma}+\frac{1}{sin\theta_1}\frac{J_1' (u)}{J_1(u)} \, ,
\label{equation_B17}
\end{equation}

where $u=k\, a\, sin\theta_1 = \kappa \, a$ .\newline

Eliminating $\gamma$ by using (\ref{equation_B16}) and (\ref{equation_B17}) it is possible to obtain, after some manipulation

\begin{equation}
\frac{y}{k \, a}= \frac{1}{u^2}\frac{J_1(u)}{u\, J_1' (u)}\left[ \left( \frac{u\, J_1' (u)}{J_1(u)}\right)^2-1+\frac{u^2}{(k\, a)^2}\right] \, .
\label{equation_B18}
\end{equation}

Since this analysis is restricted to the case of large waveguides with big aperture radius, thus $k\, a>>1$ and $\theta_1\simeq0\rightarrow cos\theta_1\simeq1$, and for finite values of $u$ and $y$ so (\ref{equation_B18}) can be simplified to 

\begin{equation}
 \left( \frac{u\, J_1' (u)}{J_1(u)}\right)^2-1=0 \; .
\label{equation_B19}
\end{equation}

The Eq. (\ref{equation_B19}) has two roots given by

\begin{equation}
\frac{u\, J_1' (u)}{J_1(u)}=\pm1 \; ,
\label{equation_B20}
\end{equation}

which corresponds to the definition of $\gamma$ given in (\ref{equation_B16}) giving $\gamma=\mp1$.\newline
\newline The well known relationship between the Bessel functions and their derivatives leads from (\ref{equation_B20}) to the next solutions

\begin{equation}
J_0(u)=0 \; 
\label{equation_B21}
\end{equation}

and

\begin{equation}
J_2(u)=0 \; .
\label{equation_B22}
\end{equation}
\newline Applying these to (\ref{equation_B18}) leads to two groups of solutions (for large values of the product $ka$)

\begin{equation}
u\simeq\left\{\begin{array}{cc}  P_{0\,l} \\P_{2\,l} \;, \end{array}\right.
\label{equation_B23}
\end{equation}

where $l=1,2,...$ . \newline
\newline Therefore, two sets of modes exist inside of the corrugated waveguide, denominated $HE_{1\,l}$ modes (given by the roots of $J_0$ ,i.e., $P_{0\,l}$) and the $HE'_{1\,l}$ modes (given by the roots of $J_2$, i.e., $P_{2\,l}$). 

It can now be shown that for $HE_{1\,l}$ modes under the conditions

\begin{equation}
\Bigg\{ \begin{array}{cc}  \gamma\rightarrow1 \\k\,a\rightarrow\infty\\u=P_{0\,l} \;, \end{array} 
\label{equation_B24}
\end{equation}
the following values of $u$ and $\gamma$ are obtained
\begin{equation}
u=u_l=P_{0\,l}\Big\{1-\frac{y}{2\,k\,a}-\frac{1}{2}\left[1-\frac{y^2}{4}(1+P_{0\,l}^2)\right]\left(\frac{1}{k\,a}\right)^2
+\frac{y}{4}\left[1-\frac{y^2}{12}(7P_{0\,l}^2+1)\right]\left(\frac{1}{k\,a}\right)^3+... \Big\}  
\;
\label{equation_B25}
\end{equation}

\begin{equation}
\gamma=1-P_{0\,l}^2\Big\{\frac{y}{2\,k\,a}-\frac{y^2}{8}(4+P_{0\,l}^2)\left(\frac{1}{k\,a}\right)^2-\frac{y}{2}\left[1-\frac{1}{2} P_{0\,l}^2-\frac{y^2}{4}(3P_{0\,l}^2+2)\right]\left(\frac{1}{k\,a}\right)^3+... \Big\} \;.
\label{equation_B26}
\end{equation} \newline

And for the $HE'_{1\,l}$ modes, under the analogous conditions to (\ref{equation_B24}) given by
\begin{equation}
\Bigg\{ \begin{array}{cc}  \gamma\rightarrow-1 \\k\,a\rightarrow\infty\\u=P_{2\,l}\;, \end{array} 
\label{equation_B27}
\end{equation}

the following values of $u$ and $\gamma$ are obtained

\begin{equation}
u=u'_l=P_{2\,l}\Big\{1-\frac{y}{2\,k\,a}+...\Big\} \label{equation_B28}
\end{equation}

\begin{equation}
\gamma=-1-P_{2\,l}^2\Big\{\frac{y}{2\,k\,a}+...\Big\} \label{equation_B29} \;.
\end{equation} \newline

The fields expressions for the transverse waves in the corrugated waveguide can be now derived from (\ref{equation_B10}) to (\ref{equation_B15}), leading to \newline

for $HE_{1\,l}$ modes

\begin{equation}
\vec{E_t}=-j\frac{k\,a}{u}A\left[J_0\left(\frac{r}{a}u\right)\vec{u}_x+\frac{1}{4}u^2\frac{y}{k\,a}J_2\left(\frac{r}{a}u\right)\Big(cos(2\phi)\vec{u}_x+sin(2\phi)\vec{u}_y\Big)\right]e^{-j\beta z} \;,
\label{equation_B30}
\end{equation}

and for $HE'_{1\,l}$ modes 

\begin{equation}
\vec{E_t}=-j\frac{k\,a}{u}A\left[J_2\left(\frac{r}{a}u\right)\Big(cos(2\phi)\vec{u}_x+sin(2\phi)\vec{u}_y\Big)\right]e^{-j\beta z} \;.
\label{equation_B31}
\end{equation} \newline

Note that (\ref{equation_B30}) and (\ref{equation_B31}) are the equations used in (\ref{equation_39}) and (\ref{equation_40}) respectively.

\end{appendices}


\newpage

\begin{thebibliography}{9}

\bibitem[Minnett \& Thomas, 1966]{1966ITAP...14..654M} Minnett, H., \& Thomas, B.\ [1966] {\it IEEE Transactions on Antennas and Propagation} {\bf 14}, 654.

\bibitem[Granet \& James, 2005]{2005ATM} Granet, C., \& James, C.\ [2005] {\it IEEE Antennas and Propagation Magazine} {\bf 47}, 2, 76-84.



\bibitem[Teniente et al., 2002]{2002MWC} Teniente, J., D. Goni  Gonzalo, R., Del Rio, C.\ [2002] {\it IEEE Microwave and Wireless Components Letters } {\bf 1}, 200-202.

\bibitem[Clarricoats \& Olver, 1984]{Clarricoats} Clarricoats, P.J.B., \& Olver, A.D.\ [1984] {\it IEE Electromagnetic Waves Series} {\bf 18}.



\bibitem[Dragone, 1977] {Dragone} Dragone, C. [1977]  {\it The Bell System Technical Journal} {\bf 56}, 835-867.

\bibitem[Pozar, 2011] {Pozar} Pozar, D. M. [2011] {\it Microwave Engineering}, 4th Ed. (University of Massachusetts at Amherst), 121-130.

\bibitem[Dragone, 1977] {Dragone2} Dragone, C. [1977]  {\it The Bell System Technical Journal} {\bf 56}, 869-888.

\bibitem[Mac \& Thomas, 1969] {Mac} Mac, B., \& Thomas, A. [1969]  {\it Electron. Lett.} {\bf 6}, No.22, 561-563.

\bibitem[Sievenpiper, 1999] {Sievenpiper} Sievenpiper, D.F. [1999] {\it High-Impedance Electromagnetic Surfaces}  (University of California Los Angeles).

\bibitem[M. Navarro, 2009] {Navarro} M. Navarro-Cía, M. Beruete, S. Agrafiotis, F. Falcone, M. Sorolla, and S. Maier, [2009] {\it Broadband spoof plasmons and subwavelength electromagnetic energy confinement on ultrathin metafilms}. {\it Opt. Express} {\bf 17}, 18184-18195.

\bibitem[Hibbins, 1999] {Hibbins} Hibinns, A.P. [1999] {\it Grating Coupling of Surface Plasmon Polaritons at Visible and Microwave Frequencies} (University of Exter).

\bibitem[Lockyear, 2004] {mlj} Lockyear, M.J. [2004] {\it Electromagnetic Surface Wave Mediated Absortion and Transmission of Radiation at Microwave Frequencies } (University of Exter).

\bibitem[Brock, 2013] {Brock} Brock, E.M.G [2013] {\it The Lateral Confinement of Microwave Surface Waves} (University of Exter).

\bibitem[Kildal, 1990] {Kildal} Kildal, P.S. [1990] {\it IEEE Transactions on Antennas and Propagation} {\bf 38}, 1537-1544.

\bibitem[Sievenpiper et al., 1999] {Sievenpiper2} Sievenpiper, D., Zhanf, L., Broas, R.~F.~J, Alexopolous, N.~G. \& Yablonovitch, E. [1999] {\it IEEE Transactions on Microwave Theory Techniques} {\bf 47}, 2059-2074.

\bibitem[Silveirinha et al., 2008] {Silveirinha} Silveirinha, M.~G., Fernandes, C.~A. \& Costa, J.~R. [2008] {\it IEEE Transactions on Antennas and Propagation} {\bf 56}, 405-415.

\bibitem[Yang et al., 1999] {Yang} Yang, F.R., Kuang-Ping, M., Yongxi, Q. \& Itoh, T. [1999] {\it IEEE Transactions on Microwave Theory and Techniques} {\bf 47}, 2092-2098.

\bibitem[Sievenpiper et al., 2003] {Sievenpiper3} Sievenpiper, D.~F., Schaffner, J.~H., Song, H.~J., Loo, R.~Y. \& Tangonan, G. [2003] {\it IEEE Transactions on Antennas and Propagation} {\bf 51}, 2713-2722.

\bibitem[Shahcheraghi \& Yahaghi, 2015] {Shahcheraghi} Shahcheraghi, S. \& Yahaghi, A. [2015] {\it Progress In Electromagnetics Research M} {\bf 44}, 109-118.

\bibitem[Scarborough et al., 2013] {Scarborough} Scarborough, C.P., Wu, Q.,  Werner, D.H., Lier, E., Shaw, R.K. \& Martin, B.G. [2013] {\it IEEE Transactions on Antennas and Propagation} {\bf 61}, 1081-1088.

\bibitem[Scarborough et al., 2014] {Scarborough2} Scarborough, C.P., Werner, D. H. \& Wolfe D. E. [2014] {\it Advanced Antennas Enabled By Electromagnetic Metamaterials} (Penn State University).

\bibitem[Lier et al., 2011] {Lier}  Lier, E., Scarborough, C.P., Wu, Q. \& Bossard, J.A. [2011] {\it Nature Materials} {\bf 10}, 216.

\bibitem[Werner, 2017] {Werner} Douglas Werner, D.H. [2017] {\it Broadband Metamaterials in Electromagnetics: Technology and Applications}, ed. Pan Stanford {\bf 382}.

\bibitem[Hoyland et al., 2012] {MFI} Hoyland, R.J., Aguiar-Gonz\'alez, M., Aja, B.,  
	{Ari{\~n}o}, J., {Artal}, E., Barreiro, R.B., Blackhurst, E.J., 
	Cagigas, J., Cano de Diego, J.L., Casas, F.J., 
	Davis, R.J., Dickinson, C., Arriaga, B.E., 
	Fernandez-Cobos, R., de la Fuente, L., G{\'e}nova-Santos, R., 
	G{\'o}mez, A., Gomez, C., G{\'o}mez-Re{\~n}asco, F., 
	Grainge, K., {Harper}, S., Herran, D., Herreros, J.M. , Herrera, G.A., Hobson, M.P., Lasenby, A.N., Lopez-Caniego, M., L\'opez-Caraballo, C., Maffei, B., 	Martinez-Gonzalez, E., McCulloch, M., Melhuish, S., Mediavilla, A., Murga, G., Ortiz, D., Piccirillo, L., 
	Pisano, G., Rebolo-L\'opez, R., Rubi\~no-Martin, J.A., Ruiz, J.L., Sanchez de la Rosa, V.´, Sanquirce, R., Vega-Moreno, A., Vielva, P., Viera-Curbelo, T., Villa, E., Vizcarg\"uenaga, A. \& Watson, R.A. [2012] {\it Proceedings of the SPIE}, {\bf 8452}.

\bibitem[Luukkonen et al., 2008] {Luukkonen}  Luukkonen, O., Simovski, C.R., Granet, G., Goussetis, G., Lioubtchenko, D.V., Raisanen, A.V. \& Tretyakov, S.A. [2008] {\it IEEE Transactions on Antennas and Propagation} {\bf 56}, 1624-1632.

\bibitem[Zeng et al., 2019] {Zeng}  Zeng, L., Bennett, C., Chuss, D.T., Wollack, E.J. [2010] {\it Proceedings of the SPIE}, {\bf 7741}.

\bibitem[Teniente et al., 2002] {Del Río}  Teniente, J and Gonzalo, Ram\'on and Del R\'io, Carlos. [2002] {\it doi:10.1109/APS.2002.1018342}.

\bibitem[Del Torto et al., 2015] {Milano}  F. Del Torto, C. Franceschet, F. Villa, P. Battaglia, M. Bersanelli, F. Cavaliere, M. Gervasi, A. Gregorio, A. Mennella, G. Morgante, O. A. Peverini, F. Pezzotta, A. Zacchei, and M. Zannoni. [2015] {\it 36th ESA Antenna Workshop on Antennas and RF Systems for Space Science}.


\end{thebibliography}
\end{document}